\documentclass{aastex}

\usepackage{amsmath}
\usepackage{natbib}
\usepackage{emulateapj5}

\newcommand{\msun}{{M}_{\odot}}
\newcommand{\lsun}{{L}_{\odot}}

\newcommand{\neon}{\ensuremath{^{22}\mathrm{Ne}\;}}
\newcommand{\Ne}{\ensuremath{^{22}\mathrm{Ne}\;}}
\newcommand{\h}{\ensuremath{\mathrm{H}\;}}
\newcommand{\he}{\ensuremath{^{4}\mathrm{He}\;}}
\newcommand{\carb}{\ensuremath{^{12}\mathrm{C}\;}}
\newcommand{\oxy}{\ensuremath{^{16}\mathrm{O}\;}}
\newcommand{\oxyet}{\ensuremath{^{18}\mathrm{O}\;}}
\newcommand{\nitr}{\ensuremath{^{14}\mathrm{N}\;}}
\newcommand{\flor}{\ensuremath{^{18}\mathrm{F}\;}}
\newcommand{\xne}{\ensuremath{X_{22}} }

\newcommand{\ee}[2]{\ensuremath{#1 \times 10^{#2}}}

\renewcommand{\vec}[1]{\boldsymbol{#1}}
\newcommand{\rhat}{\vec{\Hat{r}}}

\newcommand{\der}[2]{\ensuremath{\frac{d \,#1}{d#2}}}

\newcommand{\pder}[2]{\ensuremath{\frac{\partial \,#1}{\partial#2}}}

\shorttitle{Dynamics of \neon Settling in White Dwarf Interiors}
\shortauthors{}

\received{2002 June 26}
\begin{document}

\title{Gravitational Settling of $^{22}$Ne in Liquid White Dwarf Interiors---Cooling and Seismological Effects} 

\author{Christopher J. Deloye}
\affil{Department of Physics, University of California, Santa Barbara, Broida Hall, Santa Barbara, CA 93106}
\email{cjdeloye@physics.ucsb.edu}

\author{Lars Bildsten}
\affil{Kavli Institute for Theoretical Physics and Department of Physics, Kohn Hall,University of California, Santa Barbara, Santa Barbara, CA 93106}
\email{bildsten@kitp.ucsb.edu}

\begin{abstract}
We assess the impact of the trace element \neon on the cooling and seismology of a liquid C/O white dwarf (WD).  Due to this elements' neutron excess, it sinks towards the interior as the liquid WD cools. The subsequent gravitational energy released slows the cooling of the WD by 0.25--1.6 Gyrs by the time it has completely crystallized, depending on the WD mass and the adopted sedimentation rate.  The effects will make massive WDs or those in metal rich clusters (such as NGC 6791) appear younger than their true age.  Our diffusion calculations show that the \neon mass fraction in the crystallized core actually increases outwards.  The stability of this configuration has not yet been determined.  In the liquid state, the settled \neon enhances the internal buoyancy of the interior and changes the periods of the high radial order $g$-modes by $\approx$ 1\%.  Though a small adjustment, this level of change far exceeds the accuracy of the period measurements.  A full assessment and comparison of mode frequencies for specific WDs should help constrain the still uncertain \neon diffusion coefficient for the liquid interior.  
\end{abstract}
\keywords{diffusion---stars: abundances---stars: interiors---stars: oscillations---white dwarfs}

\section{Introduction \label{intro}}
  After \carb and \oxy, the most abundant nucleus in a $M<M_\sun$ white dwarf (WD) interior is \neon.  The reason for this starts with the physics of the CNO cycle in the hydrogen burning phase of a star's life. The slowest step in the CNO cycle is the proton capture onto \nitr and almost all of the CNO catalyst nuclei end up as \nitr at the completion of \h burning. The \nitr$(\alpha\,,\gamma)$\flor$(\beta^+)$\oxyet$(\alpha,\gamma)$\neon reaction sequence during the subsequent helium burning then processes all the \nitr into \neon.  This results in a \Ne mass fraction of $X_{22}\approx Z_{\mathrm{CNO}}\approx 0.02$ for recently formed WDs of $M<M_\sun$ (see, for example, \citet{um99}) made from stars of initial mass $\lesssim 6 \msun$.

 The possibility that \Ne could play a role in the energetics of
a cooling WD was first noted by \citet{isern91}. They discussed
the possibility of gravitational energy release if \Ne phase separates
at core crystallization. This was followed by more detailed studies
\citep{xv92, ogata93, seg94, seg96} which differed in their conclusions about the final
state once crystallization was complete; ranging from all the \Ne in
the stellar center to most of it having an unchanged profile. \citet{bild01}  (hereafter BH01) examined another aspect of \Ne which had been previously overlooked (except for a
brief mention in Bravo et al. 1992), namely its ability to sink through the
liquid WD interior.

As discussed in BH01, there is an upward pointing
electric field of magnitude $eE\approx 2m_p g$ in the degenerate WD interior, where $g$ is the local
gravitational acceleration and $m_p$ is the proton mass.  The net force on \Ne is therefore $\vec{F}=- 22m_p g \rhat + 10eE \rhat=- 2m_pg \rhat$, biasing the diffusion of \neon inward. It is the excess neutrons of the \Ne nucleus
(relative to the predominant $A=2Z$ nuclei, $A$ being the atomic mass number and $Z$ the charge) that make it special in this regard.  BH01 found that the gravitational energy released by sedimenting \neon is comparable to the thermal content of the WD at an age of a few Gyrs.  This raised the question of whether \neon could settle fast enough to significantly affect the WD's cooling significantly.  

The rate at which \neon sinks in the liquid C/O WD depends on the \neon diffusion coefficient, $D$. The conditions there are
very non-ideal, as the average Coulomb energy for the ions is
comparable to the thermal energy, or $\Gamma\equiv (Ze)^2/ a kT>1$,
where $a^3=3Am_p/4\pi \rho$. As discussed by \citet{paq86}, there is substantial (factors of many) uncertainty in the diffusion coefficients in these liquid regimes, as the familiar notions of mean free path lose their meaning.  In the absence of a definitive calculation of $D$ for this situation, BH01 proceeded by estimating $D$ for \neon in a C/O plasma by the self-diffusion coefficient, $D_s$, of the classical one-component plasma (OCP). With this, BH01 then estimated the power released, $L_g$, by \neon sinking for a fixed \neon profile. This calculation suggested that \neon sedimentation might release sufficient energy to impact WD cooling.

We now take the next step in addressing this question by performing a self-consistent evolution of both the \neon density, $\rho_{22}$, and the WD core temperature, $T_c$, in WDs composed of a single dominant ion species.  We find that \neon heating delays the time it takes a WD to cool to a given luminosity.  \emph{The total increase in cooling age by the time the WD completely crystallizes ranges from 0.25-1.6 Gyr, depending on the value of $D$  and the WD mass}. 

We also investigate the seismological impact of the \neon abundance profile at the time the WD crosses the ZZ Ceti instability strip.  The gradient in \neon abundance produces a gradient in the electron mean molecular weight, $\mu_e$, that provides an additional buoyancy and alters the Brunt-V\"{a}is\"{a}l\"{a} frequency, $N$. \emph{This contribution alters the pulsation periods of high radial order g-modes by more than the measurement errors}.  Thus $\mu_e$ gradient contributions from sedimenting \neon cannot be ignored in precision WD pulsation work such as recent contraints on the interior abundance profiles \citep{brad01} or \carb$(\alpha,\gamma)$\oxy reaction rate \citep{met01}.

An unexpected result of our calculations is the interaction between the infalling \neon and the outward moving crystal/fluid boundary as the WD cools. We assume that sedimentation halts in the crystalline interior, forcing \neon to accumulate at the crystal/fluid boundary and elevating the abundance there.  This abundance is then frozen as the crystal front moves outward.  The jump in \neon abundance at a given location depends on both the rate of \neon infall and the rate at which the crystal front moves outward.  This leads to an \neon abundance in the crystal regions of the star that \emph{increases} outward.  Whether or not this profile is subject to an elastic Rayleigh-Taylor instability remains an open question.

In $\S$  \ref{sec:methods}, we cover the details of how we model the evolution of the \neon density, $\rho_{22}$,  and the WD core temperature, $T_c$. We also discuss the current uncertainties in the OCP self-diffusion coefficient used to calculate our \neon flow rates. The details of the numeric evolution of $\rho_{22}$ are in $\S$\ref{sec:diffusion}.  In \S\ref{sec:cool} we detail the results of our self-consistent calculations. We first discuss the evolution of $\rho_{22}$, highlighting the influence of $D$, WD mass, and crystallization on the adundance profiles.  We then turn to the thermal evolution of the WD and present our new cooling curves. With the \neon abundances in hand, we estimate in \S\ref{sec:brunt} the impact \neon can have on WD pulsations, focusing on high radial order g-modes. We close in \S\ref{sec:conclusions} with a summary of our results and a discussion of unanswered questions.  The appendix discusses the Brunt-V\"{a}is\"{a}l\"{a} frequency in the deep interior.

\section{\neon Density Evolution in a Cooling White Dwarf \label{sec:methods}}

\begin{figure}
\plotone{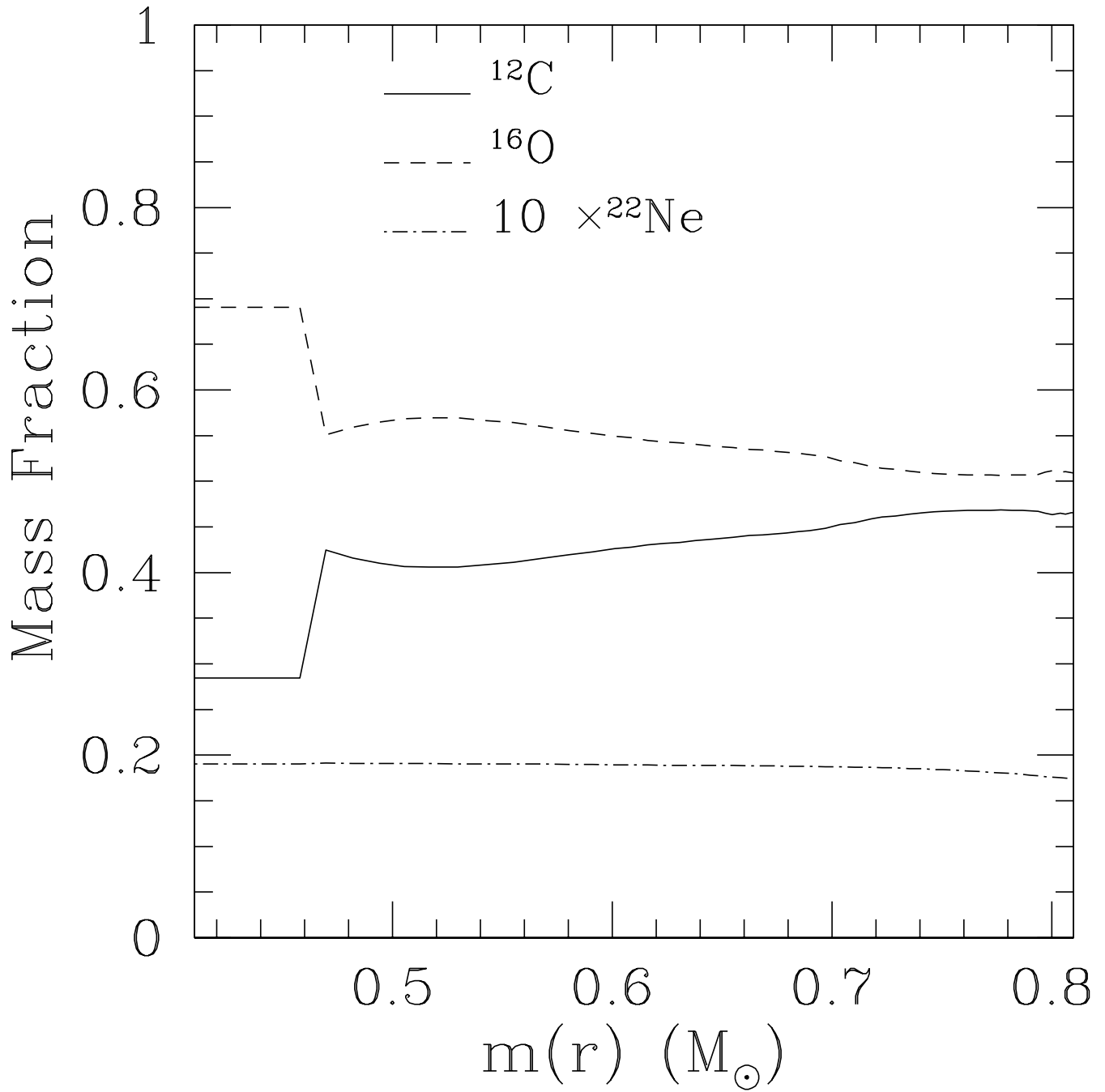}
\caption{\carb, \oxy, and \neon mass fractions of the inner $0.8 \msun$ of a $5 \msun$ star at the end of the helium burning stage(H\"{o}eflich, 2001, private communication).  Inward of 0.44 $\msun$, the mass fractions continue unchanged from their values at this point.  \label{fig:xinit}}
\end{figure}

The WD interior is a plasma of very degenerate electrons whose corresponding ions may be either in a liquid or crystalline state.  For our calculation, we presume that the WD is composed  a one component plasma (OCP) of ions whose $A$ is between 12 and 16 and whose $Z=A/2$.  We treat \neon as a trace species throughout the entire calculation, therefore neglecting its effect on the hydrostatic structure of the WD as the models evolve.  As can be seen from the initial composition profile of a realistic pre-WD model shown in Figure \ref{fig:xinit}, the \neon trace assumption is valid initially and we find that it remains valid at all times considered in our calculations. The physical state of the ions is specified by the parameter,
\begin{equation} \label{eq:gamma}
\Gamma = \frac{(Z e)^2}{a k T} = 57.7 \rho_6^{1/3} \frac{10^7 \mathrm{K}}{T} \left(\frac{Z}{8}\right)^2 \left(\frac{16}{A}\right)^{1/3}\;,
\end{equation}
where $\rho_6 = \rho/10^6$ g cm$^{-3}$.  For $\Gamma > 1$, the plasma is in a liquid state. The plasma undergoes a phase transition from the liquid to crystalline states once $\Gamma \gtrsim 173$ \citep{far93}.  

\subsection{\neon Diffusion and Sinking \label{sec:methods:neon}}
The \neon concentration evolution is governed by mass continuity
\begin{equation} \label{eq:cont}
\pder{\rho_{22}}{t} + \nabla\cdot {\bf J}_{22} =0 \;,
\end{equation}
where $\rho_{22}$ and ${\bf J}_{22}$ are the local \neon density and \neon flux respectively.  The flux has two pieces; the diffusive flux caused by
\neon density gradients and the drifting flux caused by the inward
force, $F=2m_pg$, on \neon,
\begin{equation} \label{:flux}
{\bf J}_{22} = (- D \pder{\rho_{22}}{r} - v \rho_{22}) \rhat\;,
\end{equation}
where  $D$ is the diffusion coefficient of \neon in the background plasma and $v$ is the magnitude of the local \neon drift velocity.  We assume that \neon is no longer able to sink once the WD has crystallized and fix the \neon concentration of the crystal to be that of the fluid just prior to crystallization. Thus we do not consider concentration changes due to fractionation upon freezing (see, for example, \citet{moch83, isern91}).

As mentioned in BH01, the microphysical calculation of the diffusion coefficient of \neon through a plasma that is representative of WD interior conditions has not yet been performed.  A proper treatment involves the calculation of a trace species diffusing through a multi-component plasma in both the classical and quantum liquid regimes.  The latter will be important in the higher mass WDS, where the core becomes a quantum liquid long before crystallization starts (i.e. while the core is still liquid, $\hbar \omega_p /k T > 1$; \citet{cha92}). In light of these uncertainties, BH01 estimated $D$ in two
ways. The first was using the Stokes-Einstein relation for a particle
of radius $a_p$ (taken to be the radius of the charge neutral sphere around \neon) undergoing Brownian motion in a
fluid of viscosity $\eta$, which gives $D=kT/4\pi a_p \eta$ when the
fluid is allowed to ``slip'' at the particle/fluid interface. This
estimate works well in liquids, where it is accurate
at atomic dimensions \citep{hans86}. The second was to use
the self-diffusion coefficient 
\begin{equation}
\label{eq:diff}
D_s\approx 3 \omega_p a^2 \Gamma^{-4/3},
\end{equation} 
calculated by \citet{hans75} for the OCP, 
where $\omega_p^2=4\pi n_i (Ze)^2/Am_p$ is the ion plasma frequency. These two estimates differ by 20-40 \%.  

\begin{figure}
\plotone{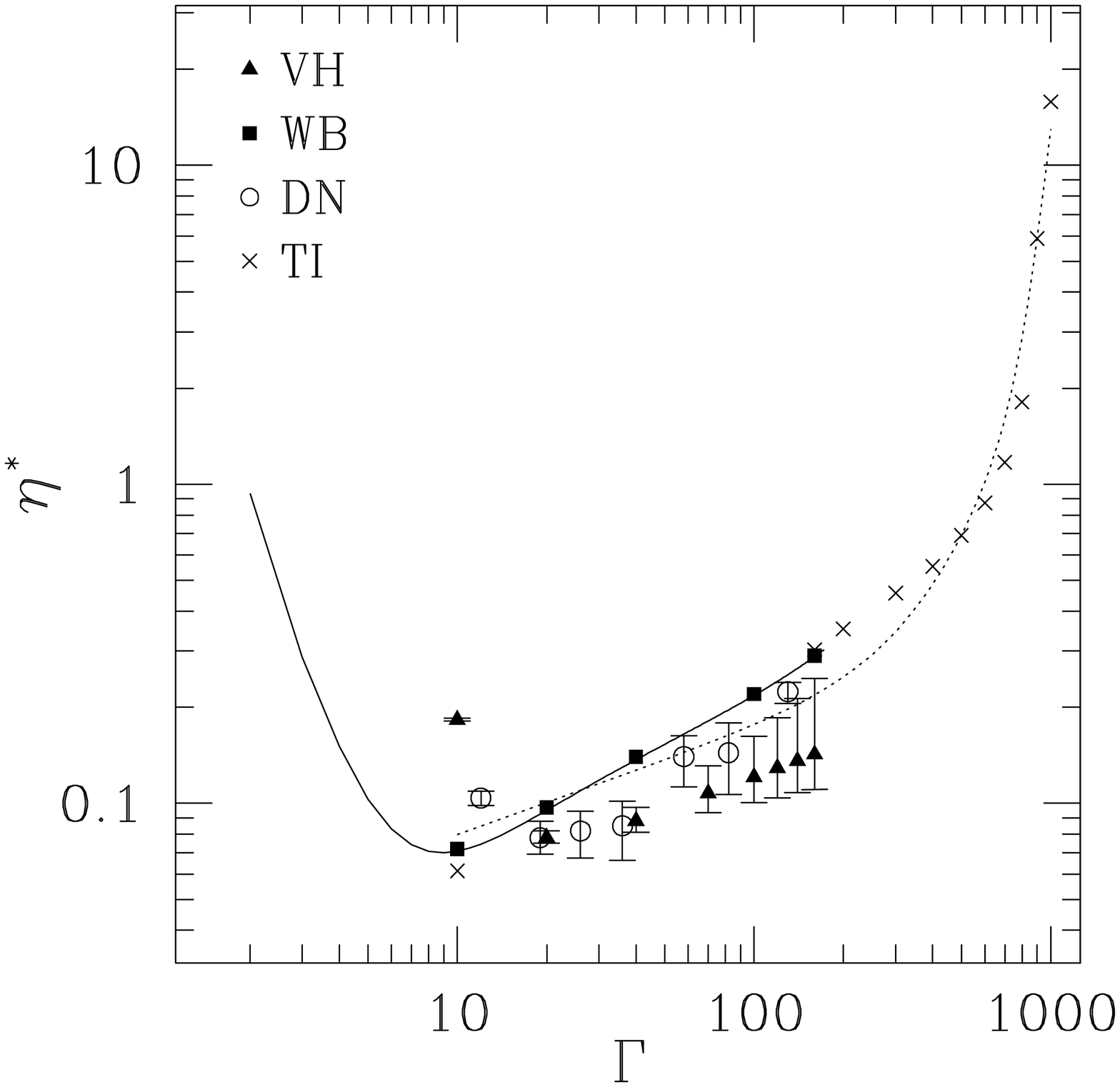}
\caption{Reduced viscosity, $\eta^* = \eta/\rho \omega_p a^2$ of the OCP. Data points are taken from: VH,\citet{vh75}; TI, \citet{ti87}; WB, \citet{wb78}; \& DN, \citet{dn00}.  Error bars are displayed for those sources which reported one.  The calculation of \citet{mur00} for the OCP limit of the Yukawa system is shown as the solid line.  The reduced viscosity inferred from the $D_s$ of BH01 merged with the fit to the data of Tanaka \& Ichimaru we utilized at high $\Gamma$ is shown as the dotted line and given as equation (\ref{eq:etaglfit}).
 \label{fig:viscosity}}
\end{figure}

In addition, there is still uncertainty in the OCP self-diffusion coefficient in the regime of interest. To highlight these uncertainties in $D_s$, we consider calculations of the OCP shear viscosity, $\eta$, which are related to the OCP self-diffusion coefficient via the Stokes-Einstein relation. A summary of shear viscosity calculations for the OCP are presented in Figure \ref{fig:viscosity}, where we show the reduced viscosity, $\eta^* = \eta/\rho \omega_p a^2$ as a function of $\Gamma$.  The results represent a range in calculation techniques---Monte Carlo \citep{vh75}, kinetic theory \citep{ti87}, and molecular dynamics \citep{dn00}.  \citet{ti87} considered a glassy state above the normal crystal transition point, thus their data extends to $\Gamma > 173$. Error bars are displayed when there was sufficient information in the literature to determine them.  Between these results, there is disagreement on the 10-30\% level. The solid line displays the analytic result of the OCP (no-screening) limit of the Yukawa system as calculated by \citet{mur00}.  The dotted line shows our fit to the data that reproduces the reduced viscosity implied from $D_s$ of equation (\ref{eq:diff}) and the Stokes-Einstein relation in the regime $\Gamma < 173$ while approximately fitting the data in the of Tanaka \& Ichimaru for $\Gamma \gg 173$.  This fit is given by
\begin{equation}
\label{eq:etaglfit}
\eta^* = 0.037 \Gamma^{0.333} + \ee{2}{-8} \Gamma^{2.7} + \ee{8}{-29} \Gamma^{9.7}\;.
\end{equation}
The first term in equation (\ref{eq:etaglfit}) is the relation between $\eta^*$ and $\Gamma$ inferred from equation (\ref{eq:diff}). We utilize this fit to explore the consequences of WD interiors not crystallizing but instead undergoing a transition to a glassy state.

With all of these uncertainties in mind, we follow BH01 and take $D= D_s$ as our fiducial best guess for the \neon diffusion coefficient.  But, as we consider the actual value of $D$ to be uncertain even in the case of \neon diffusing in a single component plasma, we carry out calculations with $D= D_s$, $5 D_s$, and $10 D_s$ to determine to what extent a reasonable variation in the diffusion rate may alter the impact of \neon sedimentation.  Although the actual $D$ may be less than $D_s$, we only consider $D>D_s$ here since for $D \ll D_s$, the impact of \neon on WD evolution is irrelevant.  The variation of up to 10 times $D_s$ is probably large for the more typical low mass WDs, but for the more massive WDs, where the interior plasma can be a quantum liquid over a substantial portion of the WD before it crystallizes, the use of the classical OCP formula may be seriously in error. 

Once $D$ is specified, the downward drift velocity, $v$, of the \neon ions under a macroscopic force $- 2 m_p g$ is
\begin{equation} \label{eq:v}
v = \frac{2 m_p g D}{k T} = \frac{18 m_p g}{Z e \Gamma^{1/3} (4 \pi \rho)^{1/2}}\;,
\end{equation}
where the latter equality holds for $D=D_s$. This is the approximation we use to determine $v$ in our calculations.  As $k T \ll U(r) \equiv \int_0^r 2 m_p g dr$ over most of the stellar radius, the motion of \neon is dominated by falling at the local drift velocity set by equation (\ref{eq:v}) (BH01).  Equation (\ref{eq:v}) also allows a rough estimate of the sedimentation timescale. The time it takes a \neon ion to sink from the surface to the center of a constant density star when it falls at the rate given by equation (\ref{eq:v}) is
\begin{equation}
\label{eq:ts}
t_s\equiv {\frac{\Gamma^{1/3}Z}{18}}\left(\frac{e^2}{G m_p^2}\right)^{1/2}
\left(\frac{1}{4\pi G \rho}\right)^{1/2}=13{\rm
Gyr} \frac{Z\Gamma^{1/3}}{6 \rho_6^{1/2}}. 
\end{equation}
(BH01).
Using the central densities as an estimate, a pure oxygen(carbon) $0.6
M_\odot$ WD at $T_c=10^7 \ {\rm K}$ has $t_s\approx 42(27) \ {\rm
Gyr}$, so that complete \Ne sedimentation is unlikely for the most
common $0.6M_\odot$ WDs.  However, the strong density dependence
means that this time decreases for massive WDs. For example, a pure
oxygen $1.2M_\odot$ WD at $10^7 {\rm K}$ has $t_s\approx 9 \,{\rm Gyr}$.
  
One caveat of our specification of the relation between $v$ and $D$ should be mentioned.  Our WD model is constructed using a fully degenerate EOS everywhere (with no inclusion of ion contributions in the structure calculation) and we take initially the \neon mass fraction to be \xne=0.02 everywhere. This choice of model and \neon profile confines us to the WD interior where 
 the electron Fermi energy $\varepsilon_F > k T$.  Modeling the outer non-degenerate parts of the star requires a more accurate background model and electric field calculation.  Since we are most interested here in the overall energetics, we can safely neglect the role of the thin outer layers.
 
\subsection{White Dwarf Thermal Evolution \label{sec:methods:thermal}}
Our WD model consists of a plasma of a single predominant ion species plus a trace amount of \neon at a local number density of $n_{22} = \rho_{22}/22 m_p$. The energetics included are the heat provided by \neon sedimentation, latent heat released upon crystallization of the major species, and the WD's radiative losses.  Since we are modeling the late ($  100$ Myr) evolution, we neglect the slight gravitational energy release as the WD contracts to its final radius.  Also, other potential heat sources such as the potential H burning in the inner envelope and fractionation effects are not included for the sake of simplicity as our goal here is to demonstrate the potential impact of \neon sedimentation separated from other sources of uncertainty in WD cooling.

The \emph{net} power generated per \neon ion by the sedimentation process is $\vec{F}\cdot\vec{v}_{bulk}$, where $\vec{v}_{bulk}$ is the bulk velocity of the \neon flow. This quantity is different from the drift velocity since we must take into account both diffusive and drifting currents.  Thus $\vec{v}_{bulk} = \vec{J}_{22}/\rho_{22}$ and the net gravitational energy release is given by
\begin{equation}
\label{eq:Lg}
L_g = \int_0^R F \frac{J_{22}}{\rho_{22}} n_{22} 4 \pi r^2 dr = \int_0^M \frac{G m(r)}{11 r^2} \xne \frac{J_{22}}{\rho_{22}} dm\,,
\end{equation}
where $n_{22} = \rho_{22}/22 m_p$.  The latent heat released upon crystallization is $l=0.77 k T$ per nuclei \citep{sal00}. For the energy lost to radiation, we use the $L-T_C$ relations of \citet{ab98}. For the heat capacity of the liquid state, $C_V$, we use equation (17) of \citet{pot00} which takes into account the ideal ion and ion-ion interactions.  We neglect the small electron contributions to the heat capacity. Quantum liquid effects are not taken into account.  In the solid state we utilize equation (5) of \citet{cha93}.

Since the thermal conduction time across the WD core is much less than the evolution time, we treat all energy sources and sinks as global quantities. The thermal evolution of the instantaneously isothermal WD at temperature $T_C$ is then 
\begin{equation}
\label{eq:Tc}
\der{T_c}{t}\,\int_0^M \frac{C_V}{A m_p} dm = -L + L_g + \frac{l}{A m_p} \der{M_{crys}}{t}\;,
\end{equation}
where $M_{crys}$ is the mass of the WD that has crystallized.  

\section{Numerical Calculation of \neon Diffusion \label{sec:diffusion}}
For a spherically symmetric flow, equation (\ref{eq:cont}) reads
\begin{equation}
\label{eq:SPcont}
\pder{\rho_{22}}{t} = \frac{1}{r^2}\pder{}{r}\left(r^2 D \pder{\rho_{22}}{r} + r^2 v \rho_{22}\right)\;.
\end{equation}
We discretize equation (\ref{eq:SPcont}) according to the prescription 
\begin{multline}
\label{eq:numcont}
\rho_i^{j+1} = \rho_i^j+ \frac{\Delta t}{{8 r_{i}^2 (\Delta r)^2}} \Big\lbrace(D_{i+1}^j+D_{i}^j) (r_{i+1}+r_{i})^2 (\rho_{i+1}^j-\rho_{i}^j)\\-(D_{i}^j+D_{i-1}^j)(r_{i}+r_{i-1})^2 (\rho_{i}^j-\rho_{i-1}^j)\\ + \frac{(v_{i+1}^j+v_{i}^j)(r_{i+1}+r_{i})^2 (\rho_{i+1}^j+\rho_{i}^j)}{2}\\ - \frac{(v_{i}^j+v_{i-1}^j)(r_{i}+r_{i-1})^2(\rho_{i}^j+\rho_{i-1}^j)}{2}\Big\rbrace\;,
\end{multline}
\citep{press92}.  Here the superscripts refer to the time index and subscripts the spatial index and the factors of two come about from the averaging of $D, r, v$ between adjacent cells. The first two terms give the diffusive fluxes through the top and bottom of the mass shell at radius $r_i$. The second two terms give the same for the drifting fluxes.  The boundary conditions are $J_{22} = 0$ at $r=0,R$, where $R$ is the WD radius.  We use $\rho_{22}$ instead of \neon concentration as our dependent variable since this simplifies the form of the continuity equation. 

The rate at which \neon flows through the interior is completely determined by $D$.  In the liquid regions of the star, we set $D = D_s$ (equation (\ref{eq:diff})).  The $\Gamma$ dependence of $D_s$ means that as the WD cools, $D$, and hence $v$, decrease and \neon flow slows. We expect \neon diffusion to halt in regions of the WD that have crystallized. There is thus an abrupt transition in the \neon flow rate at the crystal/liquid interface.  Combined with the fact that the neon concentration is fixed behind the crystal/liquid boundary, the crystallizing of the WD leaves a significant imprint on the \neon concentration profile in the star, as we discuss in \S \ref{sec:cool:diffusion}.

In order to avoid numerical difficulties, we handle the transition in the value of $D$ to zero in crystallizing regions by smoothing the transition with the formula 
\begin{equation}   
\label{eq:Dtrans}
 D_i = D_s^j \left( 1+\frac{1}{1/h+e^{(173-\Gamma)/\Delta_i}}\right)^{-1}\;.
\end{equation}
Here $1/h$ sets the factor by which $D$ is reduced from $D_s$ in the crystalline state and $2 \Delta_i$ sets the width in $\Gamma$ over which the transition from liquid to crystal occur in the $i^{th}$ shell.

By smoothing out the transition from liquid to crystal states, we effectively resolve the propagation of the crystal/liquid boundary within a single shell of our grid in the sense that it now takes a finite time for the shell to go from fully liquid to fully crystalline.  This is important because the width of the transition region affects the \neon profile near the crystal/liquid boundary.  We choose $\Delta_i$ so that the time it takes for each shell to transition from liquid to crystal (the transition regime being defined by $(1+10^{-3})^{-1}D_s \geq D \geq 10^{-3} D_s$)  equals the amount of time it takes the crystal front to move across the shell.  An easy way of approximating $\Delta_i$ is to consider the $T_c$ at which the adjacent cells in our grid crystallize.  These two temperatures set the lower and upper bounds of the range in $\Gamma$ in the $j^{th}$ between which the cell undergoes crystallization. From this $\Delta$ is determined by the simple relation
\begin{equation}
\label{eq:dgamma}
\frac{\Delta_i}{\Gamma_{crit}} \sim 1-\frac{\rho_i^{1/3}}{\rho_{i-1}^{1/3}}
\end{equation}
where $\Gamma_{crit}=173$.  For our models, the right hand side of equation (\ref{eq:dgamma}) is $\sim 10^{-3}$ and $\Delta \sim 0.1$. We presume that the latent head is released instantaneously when $\Gamma = 173$ in the shell. 

As mentioned prior, we also utilize the diffusion coefficient implied by the Stokes-Einstein relation and equation (\ref{eq:etaglfit}) to calculate the effects of \neon diffusion in WDs if the interior plasma undergoes a glassy transition instead of crystallizing. In the glassy state, we expect diffusion to continue. Thus \neon sedimentation will continue to generate energy long past the time when sedimentation would have stopped in a crystallized WD.  Obviously, such a process has the potential to maintain a WD  at a higher luminosity than it otherwise would have at very late times.  We explore this possibility in section \ref{sec:cool:cool}.

\begin{figure}
\plotone{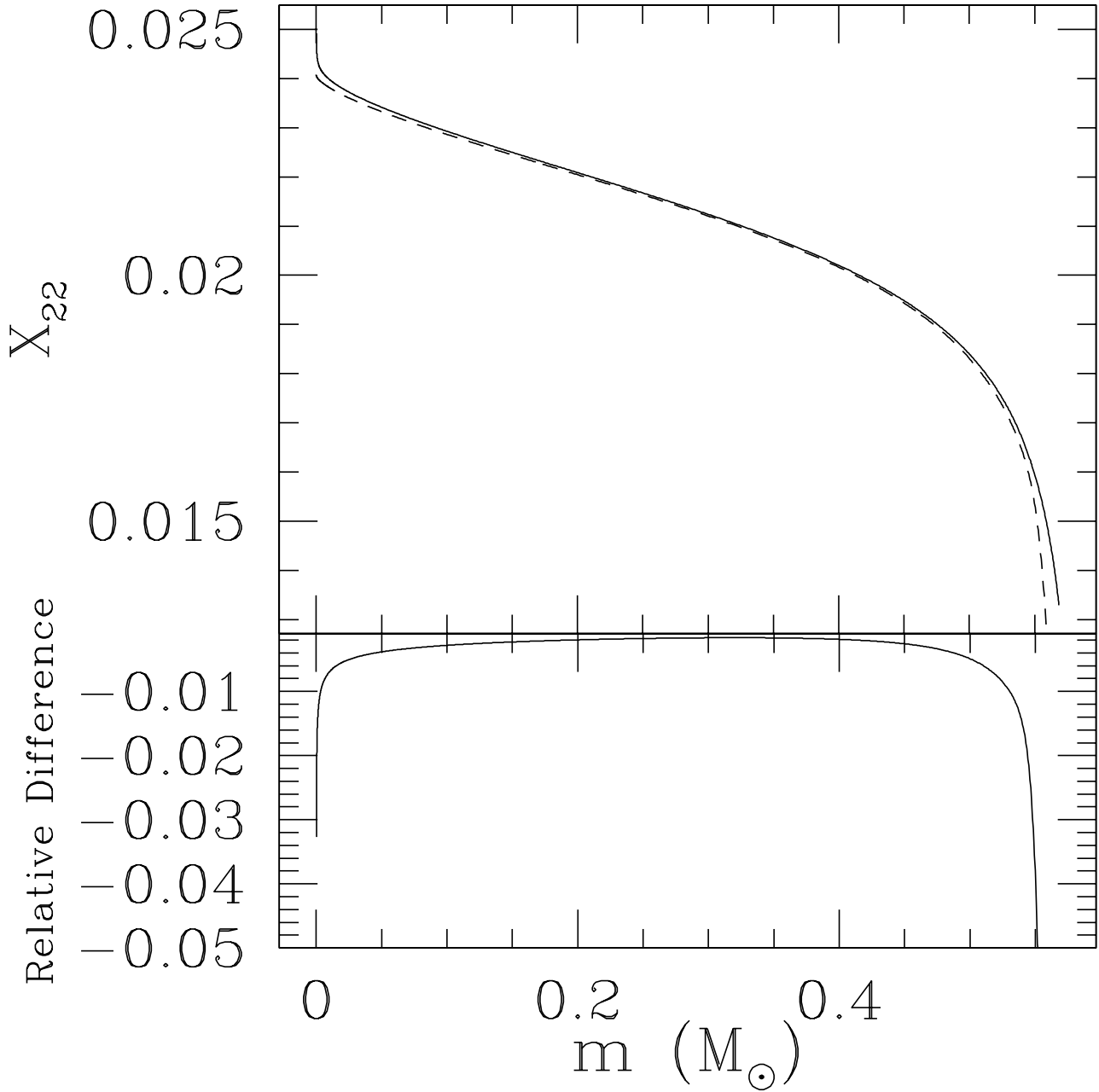}
\caption{Comparison of our differencing scheme and analytic results for a $0.6 \msun$ \nitr WD at $T_c = 10^7$ K.  The top panel shows the \neon mass fraction calculated analytically (solid lines) and numerically (dashed line or circles) at $6.3$ Gyr (left panel). The lower panel shows the relative difference in the two results (solid lines). \label{fig:xcomp1}
}
\end{figure}

Before including the thermal evolution of the WD, we performed checks on the accuracy of our algorithm at constant $T_c$ in a $0.6 \msun$ mass WD.  The motion of \neon over most of the WD is dominated by falling at the local drift speed, $v$, allowing an analytical calculation of the \neon concentration as a function of time by considering only the drift contributions to the flux.  We consider the \neon in a thin shell (at radius $r$ and thickness $\Delta r$).  From $v(r)$ we calculate the position at later times of \neon ions starting at the inner and outer boundaries of this shell.  Then, the new $\rho_{22}$ in the shell can  be calculated from the old $\rho_{22}$, $r$, and $\Delta r$ and the new $r$ and $\Delta r$.  The resulting \xne at $t=6.3$ Gyr is displayed in Figure \ref{fig:xcomp1} along with the full numerical results at the same WD age.  The lower panel of the plot displays the relative difference between the two calculations, which show a good agreement in the inner regions of the WD.

We also compared the numeric and analytic results for the diffusive equilibrium of \neon.  With $D$ and $v$ specified, the \neon profiles in diffusive equilibrium ($\vec{J}_{22}=0$) is given by
\begin{equation} \der{\ln \rho}{r} = -\frac{v}{D} = -\frac{2 m_p g}{k T}\,,\end{equation}
which gives
\begin{equation} \rho_{22eq}=\rho_{22_0} \exp\Big[\int_0^r  -\frac{2 m_p g}{k T}\; dr'\Big]\,,
\end{equation}
with $\rho_{22_0}$ normalized by total \neon mass.  A comparison between this expression and the equilibrium reached in the numeric integration shows very good agreement. 

\section{Self-Consistent \neon Diffusion and WD Cooling \label{sec:cool}}
We utilize equation (\ref{eq:numcont}) and the discretization of equation (\ref{eq:Tc})
\begin{equation}
\label{eq:numTc}
T_c^{j+1} = T_c^j+\frac{(-L^j+L_g^j) \Delta t + l^j}{C_V^j}\;,
\end{equation}
to perform the mutual evolution of $T_c$ and $\rho_{22}$ numerically. Here $L^j$, $T_C^j$,  $C_V^j$, and $l^j$ are the luminosity, central temperature, total heat capacity of the WD, and latent heat released during the $j^{th}$ timestep.  In what follows we consider a WD composed of a single ion species as the background. Figure \ref{fig:xinit} is typical of the C/O ratio in WD interiors predicted by stellar evolution models. We can approximate this composition profile by performing calculations taking the entire star to have a mean molecular weight $\mu=14$ throughout.  Therefore we consider in greatest detail the results for models with this composition.

\subsection{\neon Density Evolution in Cooling WDs \label{sec:cool:diffusion}}
In Figure \ref{fig:D1x22} we show the \neon profile evolution in the $0.6 \msun$ C/O WD when $D=D_s$ for 5 different epochs.  The initial \neon profile is a horizontal line at $X_{22} = 0.02$.  The major qualitative result of the \neon evolution is a significant depletion in the \neon concentration in the outer layers of the WD, where the rate of sinking is largest.  Elsewhere through the star, other than a gradual increase with time, the concentration is not much affected, at least as long as the WD remains in the liquid state.  This is to be expected since the sedimentation time for a low mass WD is much longer than the cooling time.

The sedimentation rate is affected by both $D$ and $M$.  We show in Figure \ref{fig:x221gyr} a comparison of the \neon profiles in our 0.6 $\msun$ C/O WD at 1 Gyr for various values of $D$.  As expected, the larger values of $D$ lead to more rapid sedimentation and a faster depletion of \neon in the outer layers of the WD and a stronger gradient in the \neon concentration as compared to $D=D_s$.  Since the sedimentation time decreases with density, \neon settles out at a faster rate in more massive WDs.  Figure \ref{fig:x22m1} shows the \neon concentrations at several ages for a 1.0 $\msun$ C/O WD.  Two effects compared to the lower mass WD are apparent. First, as expected, the rate of \neon sedimentation is greater leading to much faster depletion in the outer regions and accumulation in the central ones. But, second, the WD begins crystallizing much sooner (since the Coulomb interactions are greater) and the star finishes crystallizing at a much younger age.  This limits the extent to which \neon is able to settle inward before being frozen in place.

The depletion of \neon in the outer layers of the WD may play a role in setting the inter-diffusion properties of \carb/\oxy and \he at the core-envelope boundary. All three of these species have $A/Z\approx 2$ and the macroscopic forces acting them are approximately equal, at least in the degenerate regions of the WD.  In these regions, significant mixing of \he and the \carb/\oxy in the core should occur, at least on timescales much longer than typical diffusion times.  However, with \neon present, the $\mu_e$ of the core is slightly larger than that of the envelope and \he may be excluded from penetrating into \neon rich regions. But, as \neon is evacuated from the outer layers of the core, the buoyant barrier created by $\mu_e$ gradients between the core and envelope is removed. Thus \neon sedimentation might set a natural limit on the depth to which \he can inter-mix with the core plasma, but further work needs to be done to verify if this is truly the case. 

The time, $\Delta t$,  it takes for the \neon to evacuate the outer layers of the WD can be approximated by integrating equation (4) of BH01
\begin{equation}
\Delta t = \int_{m_1}^{m_2} \frac{Z e \Gamma^{1/3}}{18 m_p G (4 \pi \rho)^{1/2}} \frac{dm(r)}{m(r)}\,,
\end{equation}
over the outer regions of the WD, where we can take $g$ constant and the pressure to be dominated by non-relativistic, degenerate electrons.  If we make the further assumption that the layer is isothermal over the time it takes a \neon ion to fall between the limits of the integral, we find that
\begin{equation}
\label{eq:deltm}
\Delta t = 22.4 \mathrm{Gyr}\; \Gamma_m^{1/3}\, \frac{Z}{6} \left(\frac{10^8 \mathrm{cm\; s^{-2}}}{g}\right)^{0.6} \left(\frac{\Delta m}{M}\right)^{0.7}\,,
\end{equation}
where $\Gamma_m$ is the value of $\Gamma$ at the mass coordinate $m$ , $\Delta m$ is the mass outside of the depth of the ion, and $M$ is the WD mass.  Equation (\ref{eq:deltm}) is equivalent to equation (6) from BH01, but expressed in terms of mass coordinates.  Of interest is the approximate time that it takes the \neon to be depleted from the outer $10^{-2} M$ of the WD---that is from a region of roughly the same mass as the \he layer on the WD.  Taking a 50/50 mixture of \carb/\oxy in the outer core of the WD, equation (\ref{eq:deltm}) gives $\Delta t \approx 1.5 \Gamma_m^{1/3}\,\mathrm{Gyr}$, where $\Gamma_m \sim 1-10$ at these times in the outer WD .  Thus within several Gyrs, any barrier to significant \he penetration into the core should be removed.

\subsection{\neon Profile at the Crystal/Liquid Boundary \label{sec:cool:boundary}}
\begin{figure}
\plotone{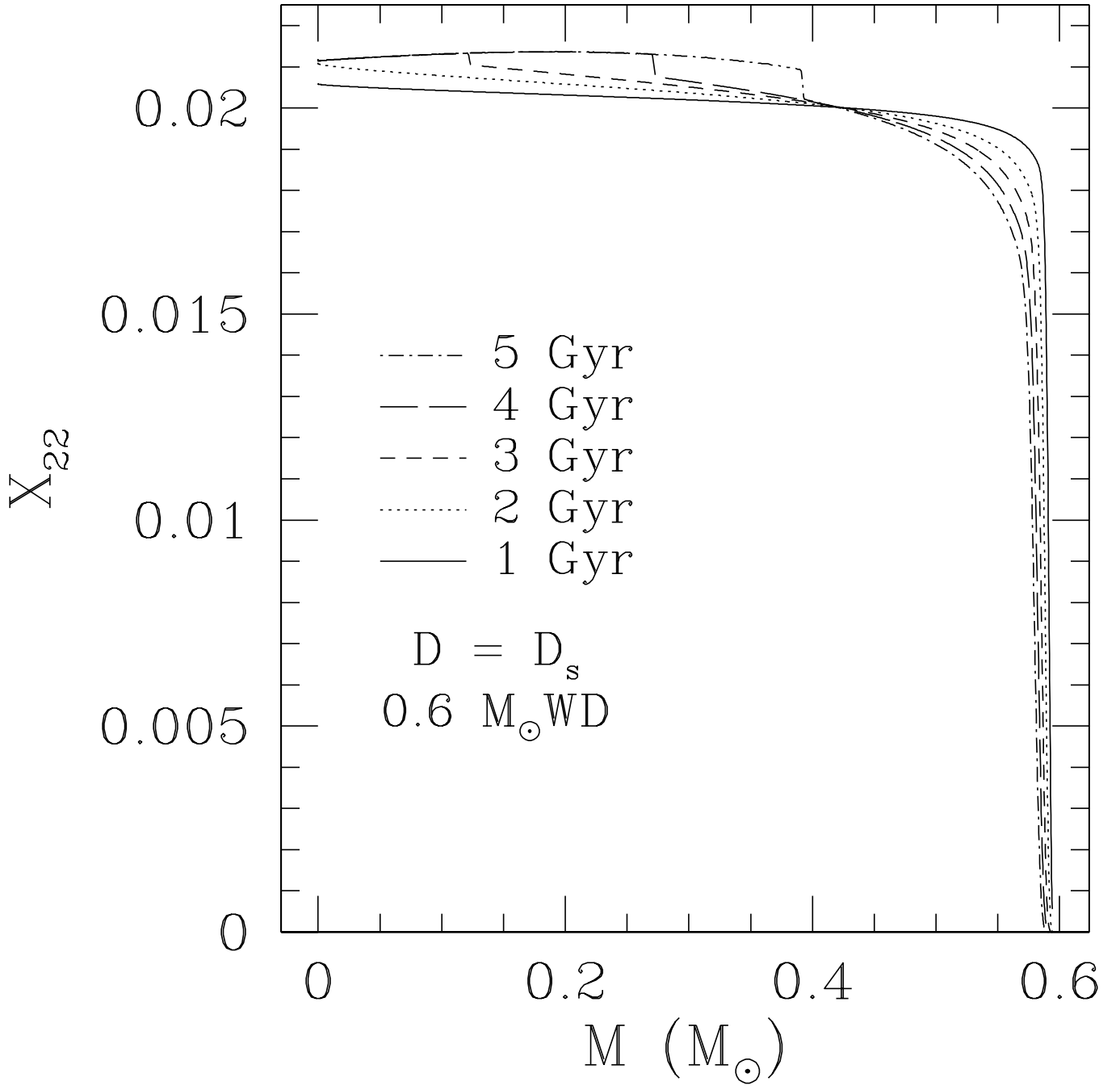}
\caption{The \neon mass fraction as a function of mass coordinate at various times in the evolution of a 0.6 $\msun$ C/O WD. \label{fig:D1x22}}
\end{figure}

\begin{figure}
\plotone{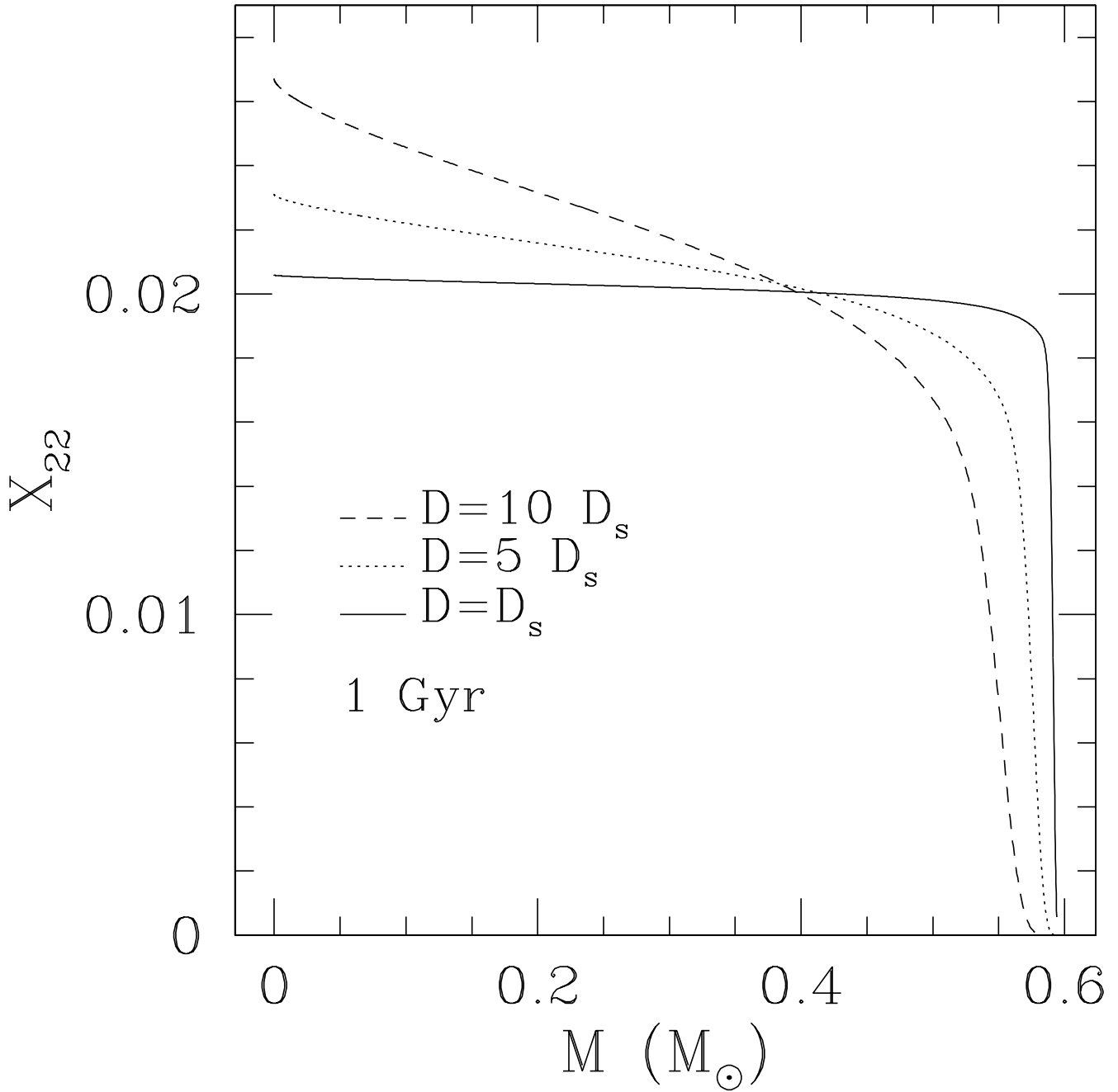}
\caption{The \neon mass fraction at 1 Gyr for the 0.6 $\msun$ C/O WD for $D = D_s, 5 D_s, 10 D_s$.  The higher sedimentation rates of the larger diffusion coefficients are apparent in the concentration profiles. \label{fig:x221gyr}}
\end{figure}

\begin{figure}
\plotone{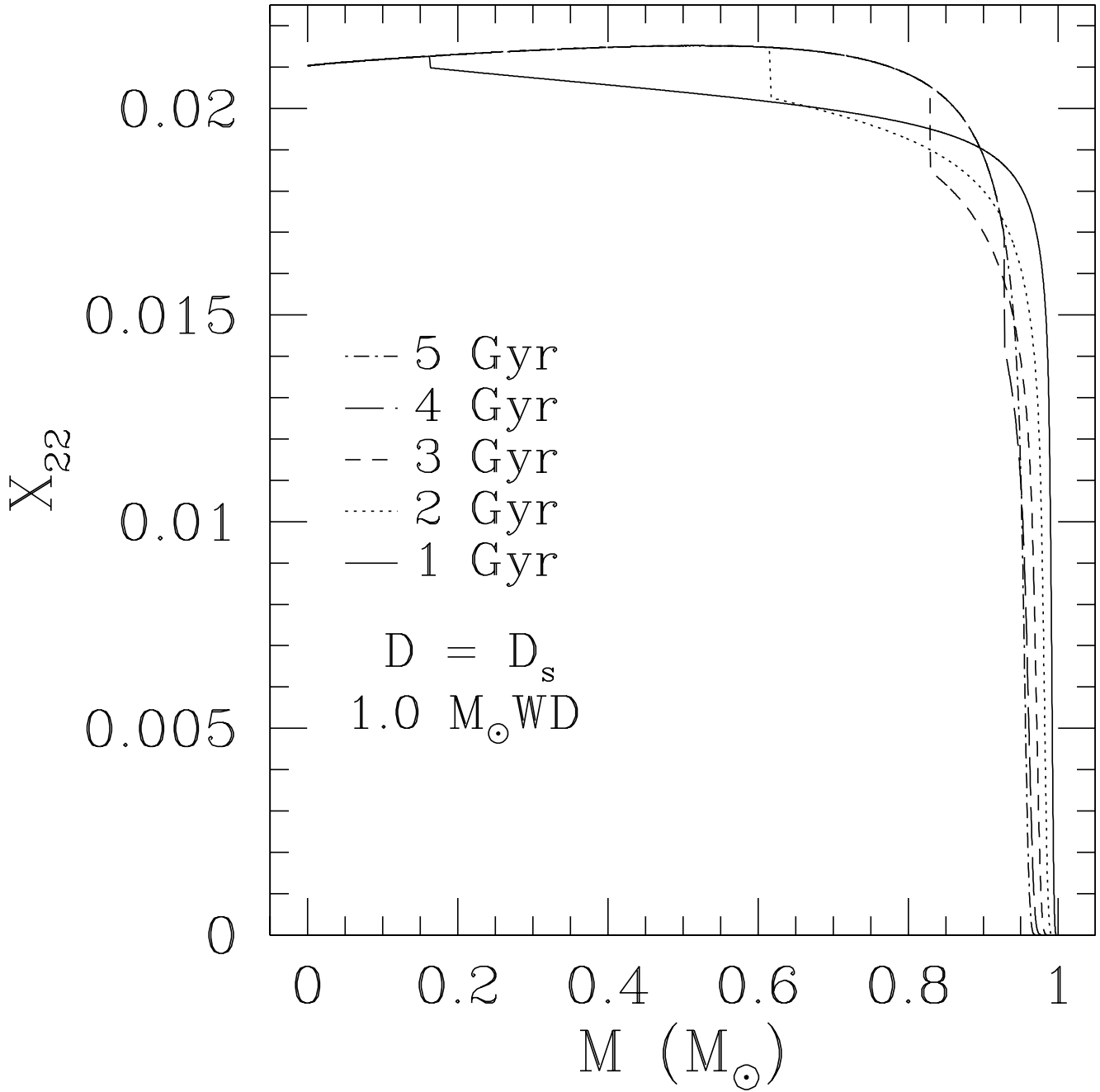}
\caption{\neon profile at several epochs in a 1.0 $\msun$ C/O WD.  Both the effects of rapid settling and rapid crystallization are seen in comparison to a 0.6 $\msun$ WD.\label{fig:x22m1}}
\end{figure}

Crystallization leaves an imprint on the \neon profile.  The location of the outward moving crystal/fluid boundary can be seen in Figure \ref{fig:D1x22} where the jump in \neon concentration in the 3 and 4 Gyr profiles occurs.  This jump is a result of the halting of \neon sedimentation in crystallized regions of the star. At the crystal/fluid interface, \neon ions falling through the liquid portion of the star encounter the impenetrable crystal boundary and accumulate.  As the star cools, the location of the crystal/fluid boundary moves outward, freezing the \neon concentration at the level to which it has accumulated.  

The question then arises as to what determines the magnitude of the jump in \neon concentration at the crystal/fluid boundary, and therefore fixes the \neon profile in the crystal portions of the WD. Consider a thin region of thickness $\Delta r$ in front of the crystal/fluid boundary, located at a radius $r_{crys}$.  At this location, the boundary position is moving outward at a rate $v_{\mathrm{crys}} = d r_{\mathrm{crys}}/dt$, the \neon falls towards the interface at a velocity $v$ and has a density of $\rho_{22}$.  In the time $t=\Delta r/v_{\mathrm{\mathrm{crys}}}$ it takes the crystal boundary to cross the region $\Delta r$, the total mass of \neon that flows into this region is
\begin{equation}m_{22} = \rho_{22} 4 \pi r_{\mathrm{crys}}^2 v t\;.\end{equation}
The \neon density in the shell between $r_{\mathrm{crys}}$ and $r_{\mathrm{crys}} + \Delta r$ after freezing is then
\begin{equation}
 \rho_{22,c} = \rho_{22} + m_{22}/(4 \pi r_{\mathrm{crys}}^2 \Delta r)\,,
\end{equation}
or defining  $\Delta \rho_{22} \equiv \rho_{22,c}- \rho_{22}$
\begin{equation}
\frac{\Delta \rho_{22}}{\rho_{22}} = \frac{v}{v_{\mathrm{crys}}}\,,
\end{equation}
since $v_{\mathrm{crys}}=\Delta r/t$. 

This analysis is valid as long as both $\rho_{22}$ and $v$ are constant over the time it takes to crystallize the shell.  Both conditions are satisfied when $\Delta \rho_{22}/\rho_{22} \ll 1$, i.e. when $v \ll v_{\mathrm{crys}}$. Equation (\ref{eq:v}) gives $v$ as a function of $\rho$, the velocity,$v_{\mathrm{crys}}$, is calculated by considering the density, $\rho_{\mathrm{crit}}$, at which $\Gamma = \Gamma_{\mathrm{crit}} = 173$ for a given $T_c$. Then $v_{\mathrm{crys}}$ is given by
\[
v_{\mathrm{crys}} = \der{r}{\rho}\Big\arrowvert_{r_{\mathrm{crys}}} \der{\rho_{crit}}{t}=-\der{r}{\rho}\Big\arrowvert_{r_{\mathrm{crys}}} \frac{9 A m_p}{4 \pi a^4} \frac{d}{d\,t}\left(\frac{(Z e)^2}{\Gamma_{\mathrm{crit}} k T}\right)\;,
\]
where $a$ is the inter-ion spacing defined in section \ref{intro}.  This evaluates to 
\begin{equation}
v_{\mathrm{crys}} = -\der{r}{\rho}\Big\arrowvert_{r_{\mathrm{crys}}} \frac{9 A m_p \Gamma_{\mathrm{crit}}^3 k^3}{4 \pi (Z e)^6} \frac{T^2 L_{\mathrm{eff}}}{C_V}\;,
\end{equation}
where $L_{\mathrm{eff}}$ represents the net rate of energy loss in the WD due to all sources and sinks and $d\,r/d\,\rho$ is calculated from the WD's $\rho(r)$.

\begin{figure}
\plotone{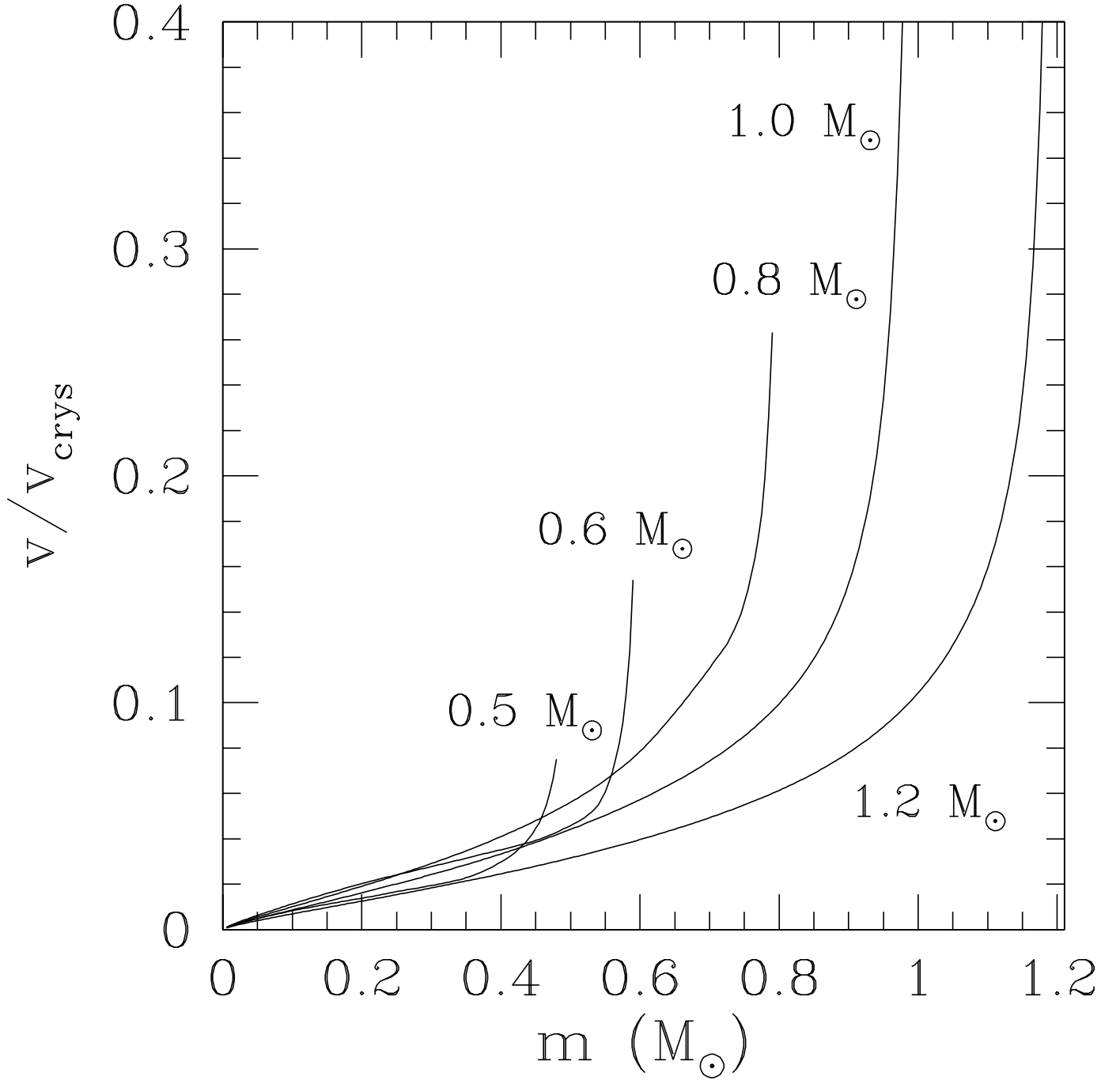}
\caption{The analytically calculated ($v/v_{\mathrm{crys}}$) for C/O WDs of 0.5, 0.6, 0.8, 1.0, and 1.2 $M_\odot$.  In regions where $v/v_{\mathrm{crys}} \ll 1$, these plots give the relative increase in \neon density at the crystal/fluid boundary. \label{fig:crysstep}}
\end{figure}

The ratio $v/v_{\mathrm{crys}}$ determines the \neon density jump across the liquid/crystal boundary.  How this ratio changes over the WD interior determines the gradient in the \neon concentration profile in the crystalline state.   A plot of $v/v_{\mathrm{crys}}$ for the models under consideration is shown in Figure \ref{fig:crysstep}.  The increase in this ratio outward is due to an increasing $v$ near the surface where $v_{\mathrm{crys}}$ decreases.  This leads to an inverted \neon concentration profile in the crystalline portion of the WD seen in Figure \ref{fig:D1x22}. This \neon abundance profile could well be destabilizing, as $\mu_e$ is increasing outwards.  However the linear stability is complicated by the elastic shear of the crystalline state, which might stabilize a profile that would otherwise be Rayleigh-Taylor unstable.  Further calculations are needed to resolve this question. 

 One may also ask the question of how the increase in \neon at the crystal/fluid boundary affects the role that fractionation plays in WD cooling.  Current thought is that under conditions typical of WDs, a C/O/\neon plasma has liquid and solid phases of equal \neon concentration \citep{seg96}.  When the WD starts to crystallize, the resulting phases are an O rich solid and C rich liquid.  The solid is the denser phase and an O rich solid core forms.  It is not until the temperature of the plasma is reduced to 1.12 times the freezing temperature of the pure carbon plasma that \neon fractionation begins to occur.  According to \citet{seg96}, by this time 70\% (by mass) of the WD has already crystallized and the additional effects of Ne distillation (see \citet{seg94}) on the cooling process is small.  If the results of \citet{seg96} (i.e. his Figure 1) are correct, the small increase in \neon concentration at the crystal/fluid boundary will not change this scenario.  To do so would require \neon concentrations of order $\xne \sim 0.1$, far above the additional increase in \xne at the crystal/fluid boundary exhibited here.  Thus, we do not expect \neon sedimentation to alter the picture developed to date of the role of \neon fractionation  in WD cooling.

We have only considered WD models of a single ion species. More realistic models with varying composition may have interesting effects on the \neon profile in the star.  For example, we expect that the viscosity of the interior plasma to increase with increasing \oxy mass fraction which produces a corresponding change in $v$.  In regions where the composition is rapidly changing, for example near 0.5 $\msun$ in Figure \ref{fig:xinit}, sharp spikes in the \neon composition may occur when the region crystallizes.  The rapid variation in composition may effect the \neon flow rates sufficiently in the liquid state to cause local increases in \neon concentrations at this stage also.  Another way in which the \neon profile could be altered is through fractionation of the C/O/\neon plasma when crystallization occurs, an effect that we have ignored in this paper.  A full calculation including \neon sedimentation and full consideration of the 3 component plasma phase transition is needed to answer the question of how much this may change the final \neon profile in the WD.

\subsection{Effects of \neon Sedimentation on WD Cooling \label{sec:cool:cool}}
\begin{figure}
\plotone{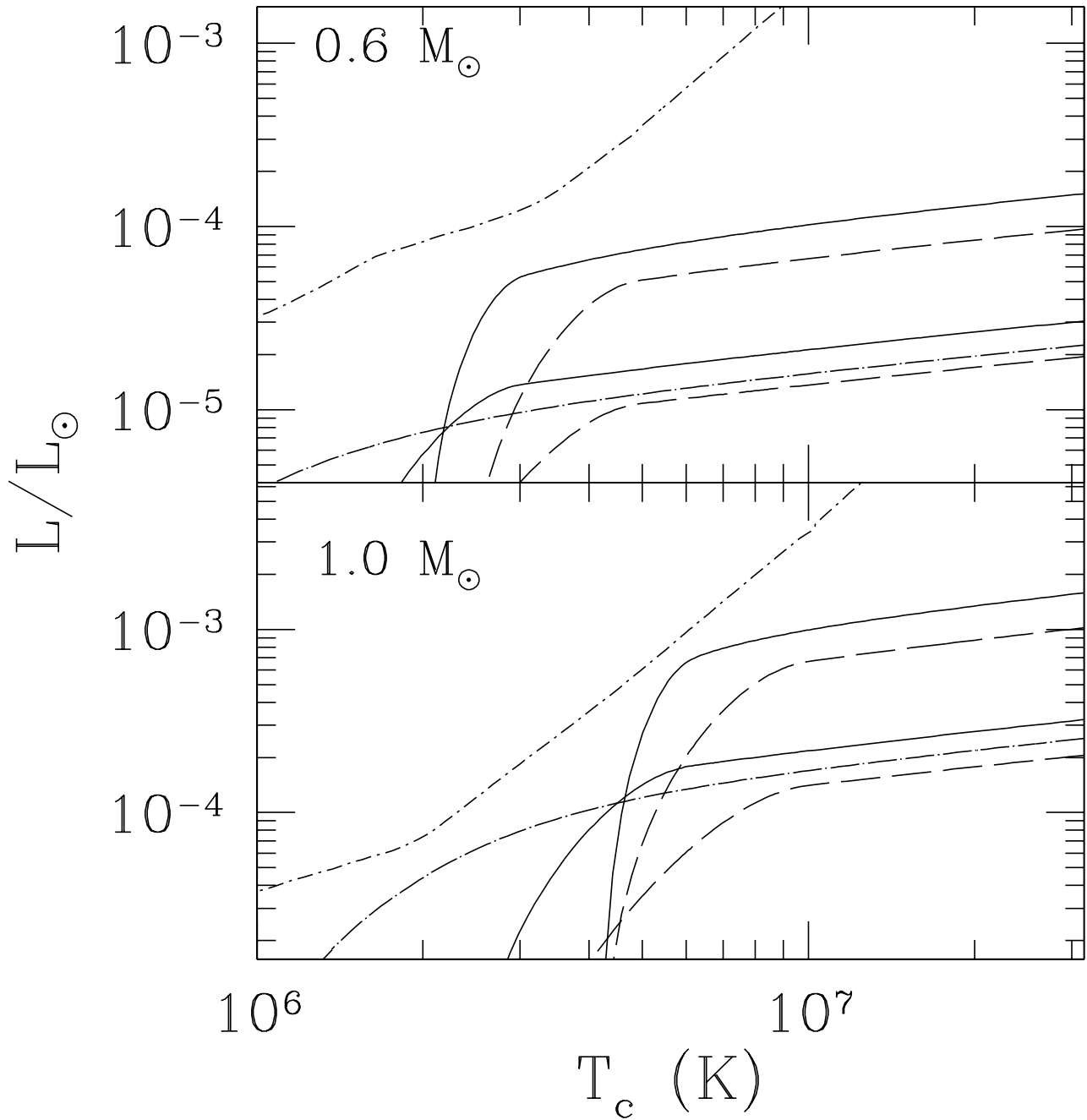}
\caption{The luminosity from \neon settling in a
$0.6M_\odot$ (top panel) and $1.0M_\odot$ (bottom panel) WD with
$X_{22}=0.02$ as a function of the WD core temperature. The solid
(dashed) lines are for a WD of pure carbon (oxygen) and show the
dependence of $L_g$ on the core composition. 
The dot-dashed lines are the $T_c-L$
relations for cooling DA WD's from \citet{ab98}. The
lower(upper) set of $L_g$ are for $D=D_s (D= 5 D_s)$. The dash-dot line shows $L_g$ in a C/O WD that undergoes a transition to a glassy state instead of crystallizing. \label{fig:TcLgcrys}}
\end{figure}

\begin{figure}
\plotone{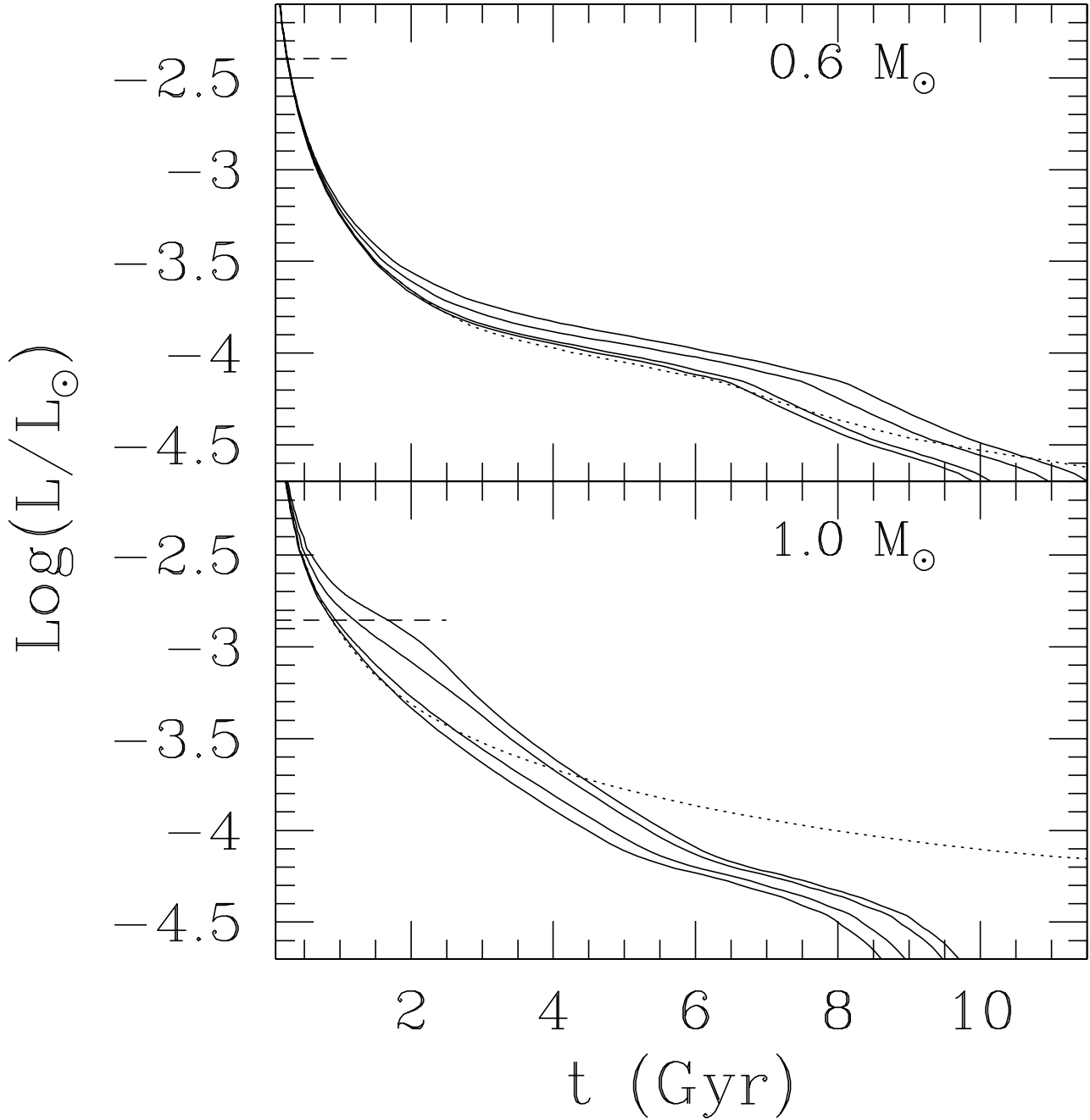}
\caption{WD luminosity as a function of time.  Both plots are for C/O WDs. The lower curve in each gives the cooling curve neglecting \neon sedimentation in the thermal evolution of the WD.  The remaining three solid lines are (from top to bottom) the results for $D=D_s$,$D=5 D_s$, and $D=10 D_s$ while the dotted line is for a model that enters a glassy state.  The glassy transition models stay more luminous at late times due to the continued sedimentation of \neon. The horizontal dashed line approximates the ZZ Ceti instability strip. \label{fig:Ltcrys}}
\end{figure}

The effects of \neon sedimentation on WD cooling are summarized in Figures \ref{fig:TcLgcrys} and \ref{fig:Ltcrys}. Figure \ref{fig:TcLgcrys} shows the instantaneous power generated by \neon sedimentation, $L_g$,  in comparison with the luminosity of the WD (dash-dot line) as a function of core temperature.  Shown are the $L_g$ produced for pure \carb and pure \oxy (solid and dashed lines respectively) stars as well as a comparison between calculations performed with $D=D_s$ and $D=5 D_s$ (lower and upper sets of curves). The results for a C/O WD are between the \carb and \oxy WD curves. The effect of crystallization in halting the \neon sedimentation is seen in the sharp change in slope for the $L_g$ curve near $T = \ee{3}{6}$ K.  Since $L_g < L$, it is seen that this WD always cools.  Thus, in the case of low mass WDs, the possibility for \neon sedimentation to provide a WD's luminosity seems unlikely, at least if WDs crystallize as they cool.

In addition, the results for a C/O WD that has a phase change to a glassy state at high $\Gamma$ values is shown for both masses by the dash-dot line.  As mentioned before, $D$ in this case was calculated from equation (\ref{eq:etaglfit}) and the Stokes-Einstein relation. That sedimentation continues to generate energy at late time is quite obvious, and in the case of a $\msun$ WD can dramatically affect the WD's late evolution.   

Concentrating on the 0.6 $\msun$, $D=D_s$ curves for a moment,  $L_g$ is substantially less than $L$ over most of the evolution of the WD and the effects of \neon sedimentation on WD cooling take some time to become apparent. This is seen in Figure \ref{fig:Ltcrys} where a comparison between the cooling with and without \neon settling is shown.  Not until after about 4 Gyr is there a substantial difference in the cooling age of this model.  The fact that the low mass WDs crystallize slowly combined with slow \neon sedimentation cause the delays from \neon sedimentation to increase until the WD is completely crystalline (which occurs at about $\log(L/L_\odot) = -4.5 $ in this model).  The total delay in cooling that occurs for this model is about 0.25 Gyr by 10 Gyr.  We also indicate the approximate location of the ZZ Ceti instability strip in this figure by the horizontal dashed line.  The \neon profile at the time at which the WD crosses this location is needed to determine the potential effects that \neon sedimentation may have on the seismological properties of the star (see \S \ref{sec:brunt}).

 As found by BH01, the competition between the heating due to \neon and the energy output of the WD is much closer in the more massive WDs, as exemplified by the 1.0 $\msun$ model in the lower panel of Figure \ref{fig:TcLgcrys}. However, in all cases we examined (including calculations with $D=10 D_s$ in the 1.0 $\msun$ model), the power generated by \neon sedimentation never exceeded the luminosity of the WD.  Thus it appears improbable in all but maybe the most massive WDs that \neon sedimentation would be able to power the WD.  However, the early competition between the \neon heating and radiative losses does show up in the cooling curves as will be seen in a moment.

The overall impact of \neon on the cooling of the more massive WDs due to this rapid sedimentation is mitigated by the equally rapid crystallization in these stars, as exhibited by the earlier turnover in the $L_g$ curves in these models. We can also see a qualitative differences in the effect of \neon sedimentation on differing mass models from this plot.  In the lower mass WDs, \neon sedimentation is a slower, enduring process which takes substantial time to compete with cooling at a significant level.  But, $L_g$ in this case stays competitive with $L$ for a longer period of time.  For the higher mass stars, sedimentation is fast and $L_g$ is large initially but drops off quickly.

In the 1.0 $\msun$ model, the $D=10 D_s$ exhibits the effects of rapid \neon depletion in the outer layers of the WD. The extremely rapid sedimentation in this model leads to a large delay in the cooling time early on as compared to the other models which include sedimentation effects.  But, as the \neon is depleted from the outer layers in this model, energy generation due to sedimentation decreases compared to the models with smaller diffusion coefficients and the relative delays between them decrease at late times.

The effects of these differences are seen in the cooling curves shown in Figure \ref{fig:Ltcrys}.  For clarity, data is only shown for WDs with $\mu=14$; results for pure \carb and \oxy WDs bracket these results.  Both plots show the effects of varying $D$ over the range of $1-10 D_s$ along with a model that neglects $L_g$ (the bottom curve in each) and a model that undergoes a transition to a glassy state instead of crystallizing (dotted curve).

\begin{figure}
\plotone{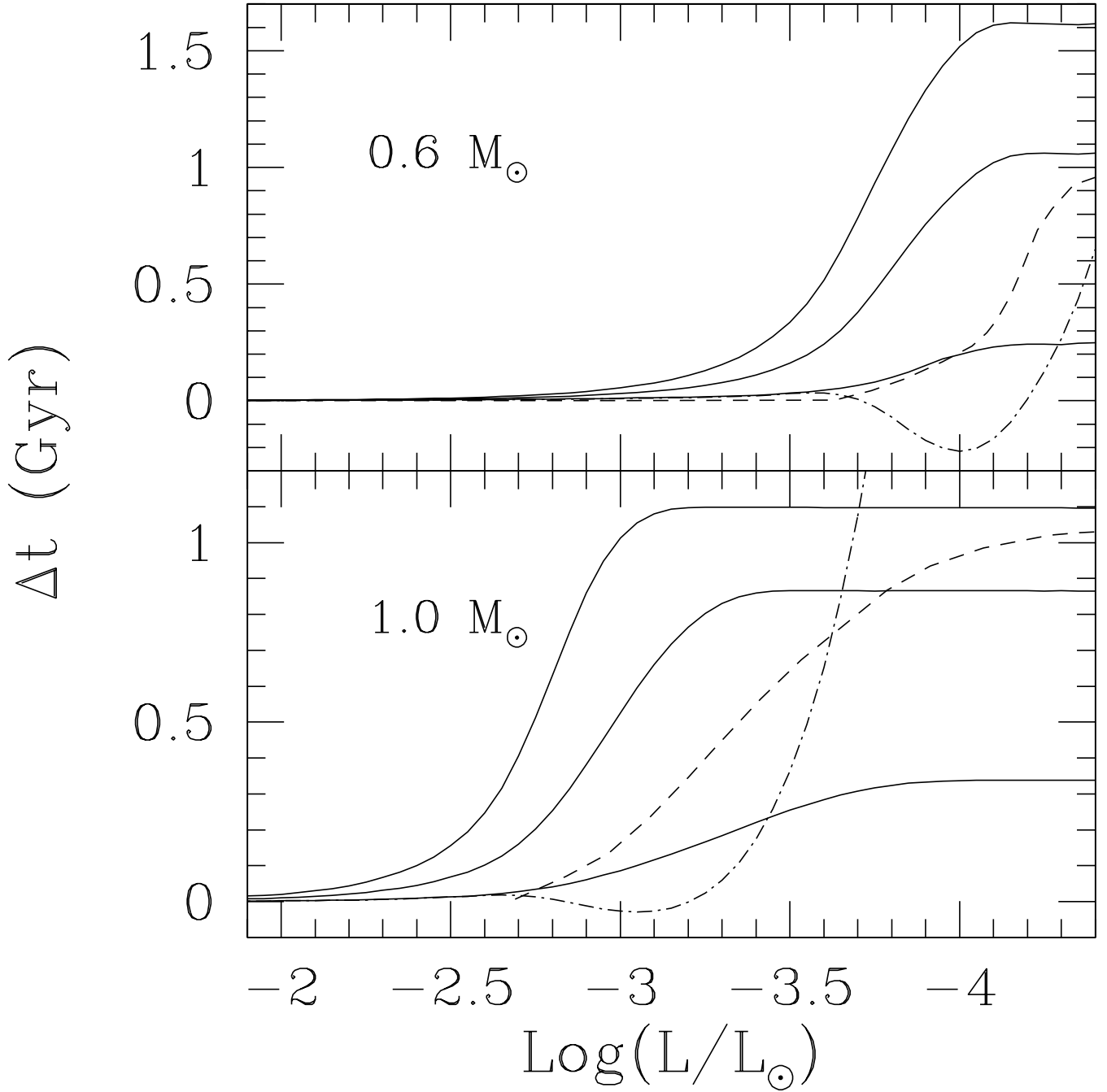}
\caption{The difference in cooling age between C/O WD models that include \neon sedimentation and a reference model that does not.  The solid lines (from top to bottom) are this age difference in models with $D=10,5,1 D_s$.  The dash-dot lines show the age difference of the glassy C/O WD model and the dash line the delays introduced by C/O fractionation as calculated by \citet{sal00} for comparison.  \label{fig:delays}}
\end{figure}

As could be expected from Figure \ref{fig:TcLgcrys}, the delays introduced in the $0.6 \msun$ model only grow significant at late times, whereas those of the $1.0 \msun$ model are apparent early.  These delays continue until the WDs have almost fully crystallized, which occurs by the time  $\log(L/\lsun) \approx -4.2 (-3.8)$ for the $0.6 (1.0) \msun$ models.  A summary of the increase in cooling ages, $\Delta t$, for these models is in Figure \ref{fig:delays}. Each panel shows the difference between the age of our $\mu=14$ models with $D = D_s, 5 D_s, 10 D_s$ (solid lines, bottom to top, respectively) and one without \neon diffusion included.  In the 1.0 $\msun$ model, the early termination of increases in $\Delta t$ in the $D = 5 D_s$ and $D=10 D_s$ models as compared to the $D=D_s$ model is the results of rapid depletion of \neon in the outer layers of the WD. So, there is little \neon left to sediment out in the remaining liquid regions of the WD before the WD completely crystallizes.

The dash-dot line in Figure \ref{fig:delays} shows the difference in age between the reference model and a WD that undergoes a glassy transition.  For a period while the reference model is undergoing crystallization, the glassy WD model is actually \emph{younger} than the reference model.  The explanation for this is that the glassy WD does not liberate latent heat, which is a much more significant quantity of energy while the reference star is crystallizing. Once the crystallization process is finished, the continued release of energy from \neon sedimentation in the glassy WD produces the age increases seen (at least in combination with the rapid cooling that the crystallized WD undergoes once it enters the Debye cooling regime). In the glassy WD models, the increase in WD age for a given luminosity continue to increase indefinitely and would lead to an abundance of faint WDs greater than would otherwise be expected.

\subsection{Cooling WDs in Metal Rich Clusters \label{sec:cool:highz}}

\begin{figure}
\plotone{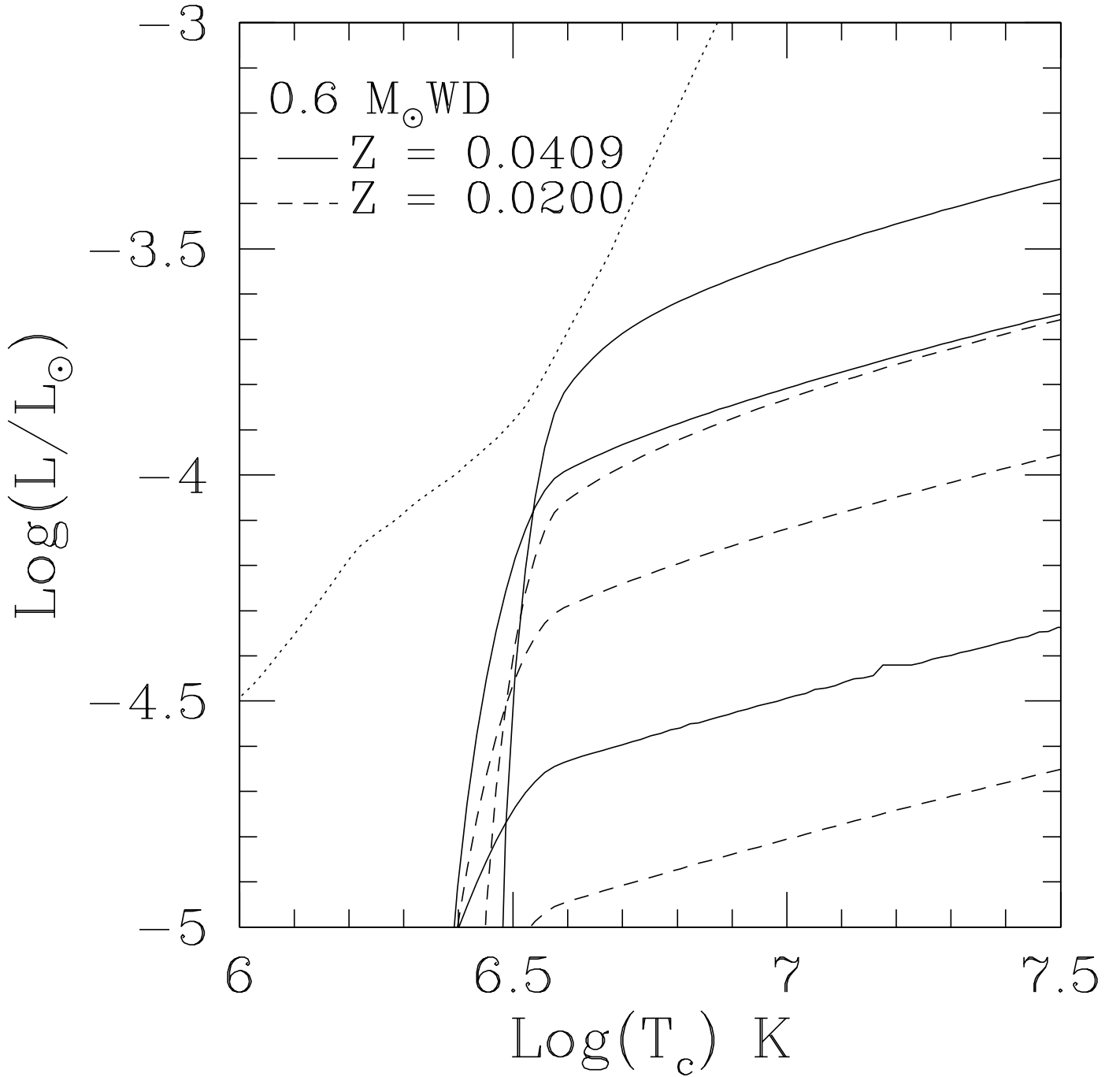}
\caption{The power generated by \neon sedimentation within WDs produced from high metallicity progenitors.  Here the solid lines give $L_g$ as a function of $T_c$ for a 0.6 $\msun$ C/O WD with a $Z = .0409$, the best fit value for NGC6791 \citep{chab99}.  For comparison, the solar metallicity results are shown as dashed lines.  The three sets show the results for $D = D_s, 5 D_s, 10 D_s$ from bottom to top. \label{fig:06LTcZcomp}}
\end{figure}

\begin{figure}
\plotone{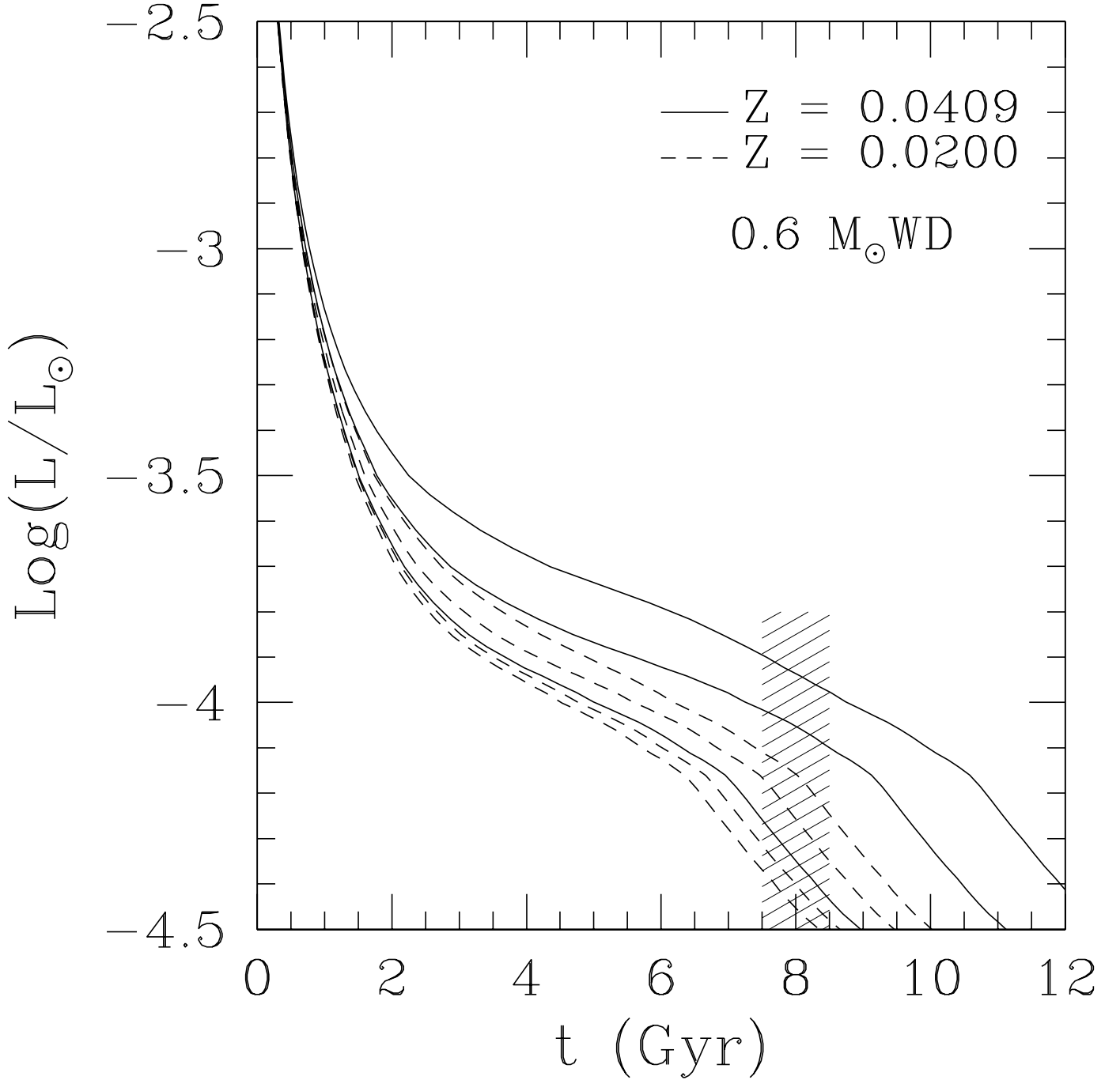}
\caption{Comparison of the cooling curves of 0.6 $\msun$ C/O WDs with progenitors of differing metallicity.  The solid curves show the results for $D = D_s, 5 D_s, \& 10 D_s$ for a WD within the cluster NGC 6791.  The dashed curves are the same for WDs whose progenitors had solar compositions, but include a model with $D=0$ (lower dashed curve) for reference. The shaded region indicates the age of NGC 6791 as determined by \citet{chab99}.  The increase in cooling ages for WDs with higher \neon content is apparent.\label{fig:06LtZcomp}}
 \end{figure}

From equation (\ref{eq:Lg}), the power generated by \neon sedimentation is proportional to \neon number density.  The delays in cooling time produced by \neon sedimentation will thus vary with the metallicity of the WD progenitor.  A particular example where this may be observationally relevant is the open cluster NGC 6791.  NGC 6791 is interesting both because is it old ($8 \pm 0.5$ Gyr) and metal \emph{rich}, $[\mathrm{Fe/H}]=0.4$ \citep{chab99}.  Figure \ref{fig:06LTcZcomp} shows a comparison in the $L_g$ generated in the 0.6 $\msun$ $\mu=14$ WD models whose progenitor's metallicity were solar (as in the prior calculations) or $Z=0.0409$, as is appropriate for WDs within NGC6791.  Shown in Figure \ref{fig:06LtZcomp} is the comparison of cooling curves between these WDs.  As is readily seen, WDs whose progenitors were metal rich will be more affected by the effects of \neon sedimentation.  This will be especially true if $D$ is significantly larger than our fiducial best guess of $D_s$. 

\begin{figure}
\plotone{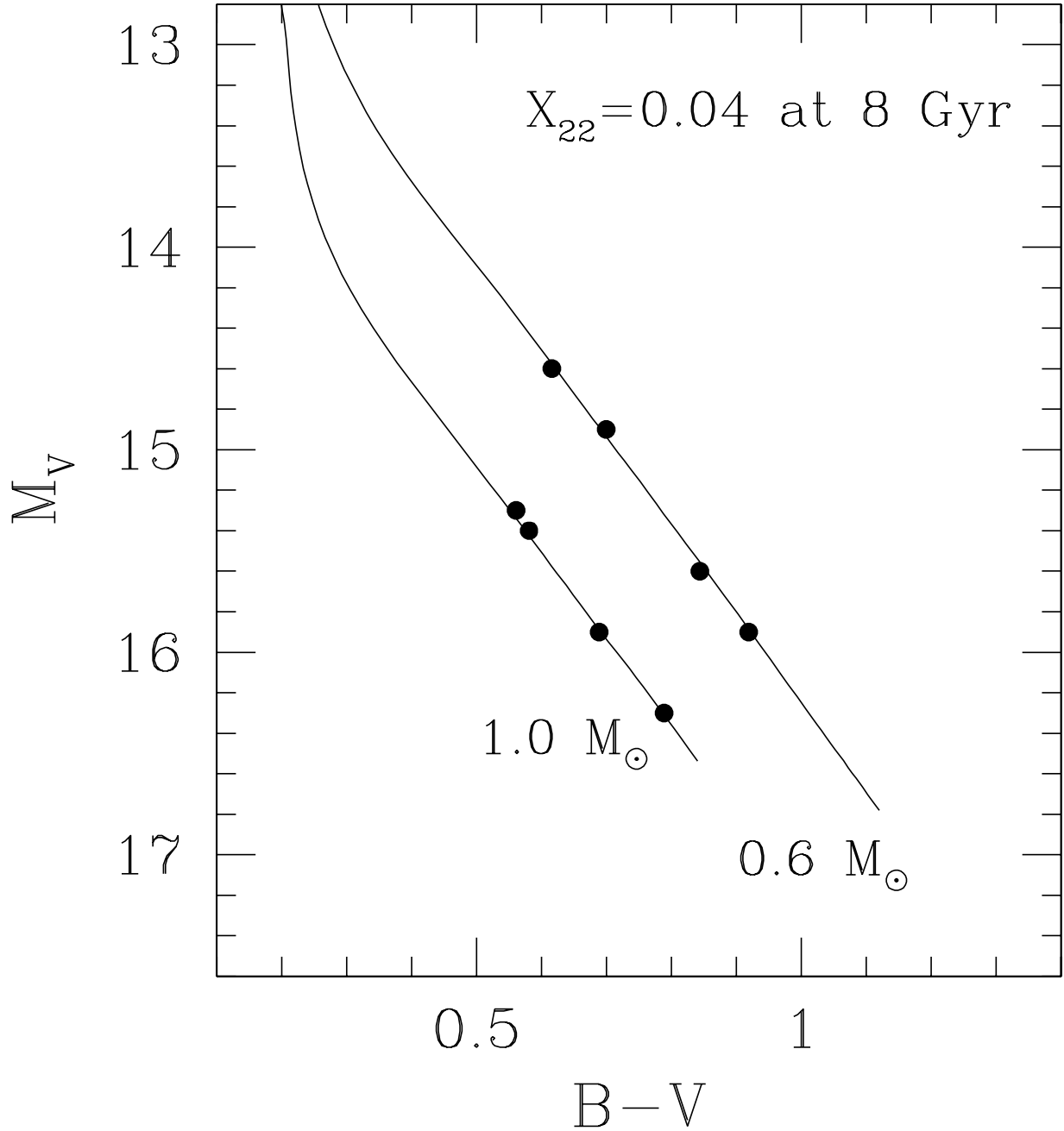}
\caption{The expected location of WDs in NGC 6791 today based on our cooling tracks with differing values of $D$.  The solid lines show the cooling tracks for the 0.6 and 1.0 $\msun$ WD models and the dots the location of these mass WDs today if $D=0\,,D_s\,,5 D_s\,,\& 10 D_s$ (from bottom to top).\label{fig:ngc6791cmd}}
\end{figure}

Can this seen?  Most likely, yes, if a massive ($M \sim \msun$) is found, as these should be equal in age to the cluster since their massive progenitors had such short lives.  Figure \ref{fig:ngc6791cmd} shows the location of a $1 \msun$ WD in the cluster CMD for different values of $D$ but all at a fixed age of 8 Gyr.  The deceleration of the aging process is evident, as the $D=D_s$ model is $\approx 0.4$ mags brighter than the model without diffusion included.  For comparison the effects on the more typical lower mass WDs are also shown.  

The fact that \neon sedimentation's effects on the WD cooling sequence is metallicity dependent also provides the most straightforward means of separating it from other uncertainties in WD cooling. To our knowledge, no other uncertainty has as significant dependence on the metal content of the WD as \neon sedimentation.  The only other effect that might is the fractionation of \neon during crystallization.  But, according to \citet{seg96}, the extent to which this effect matters is much smaller than that due to C/O fractionation and in any case occurs only after the larger portion of the WD has crystallized. 

 Separating the delays in WD cooling due to \neon from other contributions is difficult in the absence of observational comparisons between metal rich and metal poor WDs. In the low mass WDs, at the point where \neon's effects become apparent, the star has already begun crystallizing and the uncertain energy releases from fractionation can occur.  In the high mass stars, especially for large \neon diffusion rates, there is a window between 1-4 Gyrs in which \neon sedimentation can have an extremely significant impact on the cooling age of the WD.  To be fair, here again, crystallization has already started in these stars, but the potential magnitude of \neon contributions may allow separation of \neon and fractionation effects.

Although not yet feasible, potentially the clearest test of \neon sedimentation's impact on WD cooling in particular, and WD cooling theory in general, would be the construction of empirical cooling curves for WDs of specified masses.  As the overall morphology of the cooling curves can depend on which effects are included and their magnitudes, this may provide at least a qualitative means of determining empirically which phenomenon are operative in WDs.  Such a comparison would require a large sample of confirmed WDs homogeneous in mass and metallicity and whose cooling ages are determined independently of their luminosity.  
\section{Enhanced Internal Buoyancy from the \Ne Abundance Profile \label{sec:brunt}}

\begin{figure}
\plotone{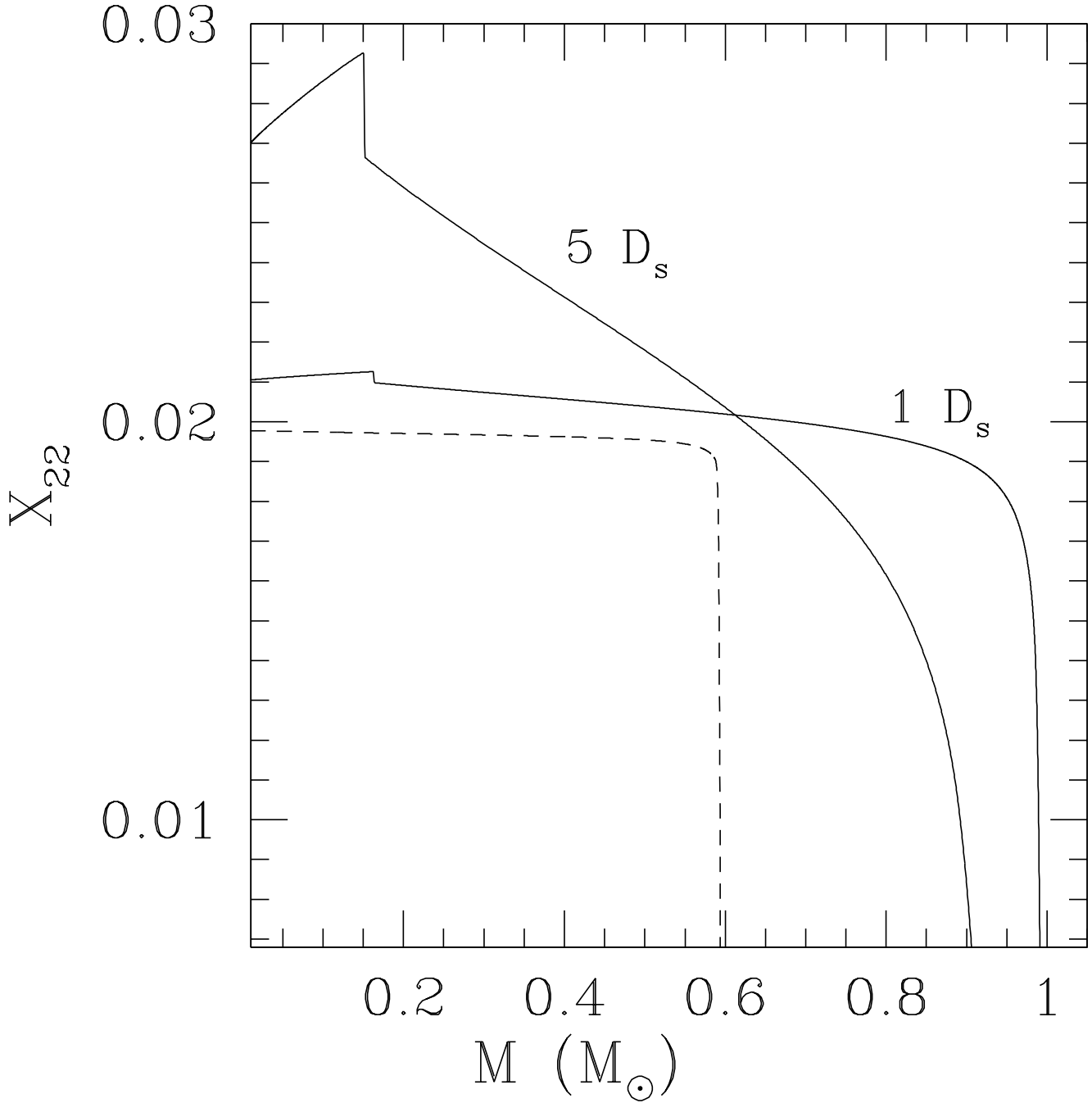}
\caption{\neon mass fraction as a function of mass for the $\mu=14$ models at the time they reach the ZZ Ceti instability strip. The dashed line is the $0.6 \msun$ model with $D= D_s$.  The solid lines give the results for a $1.0 \msun$ WD with the indicated $D$. \label{fig:neprofiles}}
\end{figure}

Typical \neon abundance profiles at the time our WD models reach the ZZ Ceti instability strip are shown in Figure \ref{fig:neprofiles}. The abundance gradient in \neon creates a gradient in the
mean molecular weight per electron, $\mu_e$, that provides an extra
restoring force for the g-modes. We now show that this modifies the
pulsation periods at levels comparable to  the uncertainties of the
measurements. To see
this, we start with the Brunt-V\"{a}is\"{a}l\"{a} frequency
\begin{equation}
N^2 = -g \left[ \der{\ln \rho}{z} - \frac{1}{\Gamma_1} \der{\ln P}{z}
\right] \;, 
\end{equation}
\citep{unno89} where $\Gamma_1= d\,\ln \rho/d\,\ln P \arrowvert_S$ is one of the adiabatic indices.  For an
isothermal background of degenerate, relativistic electrons, and
an ideal ion gas (the Coulomb corrections to $N^2$
are small; \citet{bild95}) and a $\mu_e$ gradient that
depends only on the gradients of the trace \Ne, we find (see Appendix)
\begin{equation} \label{eq:n2rel}
N^2 = \frac{3}{8}\frac{k T}{h^2 \mu_i m_p} - g \frac{1}{11} \der{X_{22}}{z} \;,
\end{equation}
where $h=P/\rho g$ and $\mu_i$ is the ion mean molecular weight.  The
first piece is the thermal contribution to $N^2$ (previously known to
be small in the deep interior; \citet{bras91}) while the second
is the \neon contribution. 

\begin{figure}
\plotone{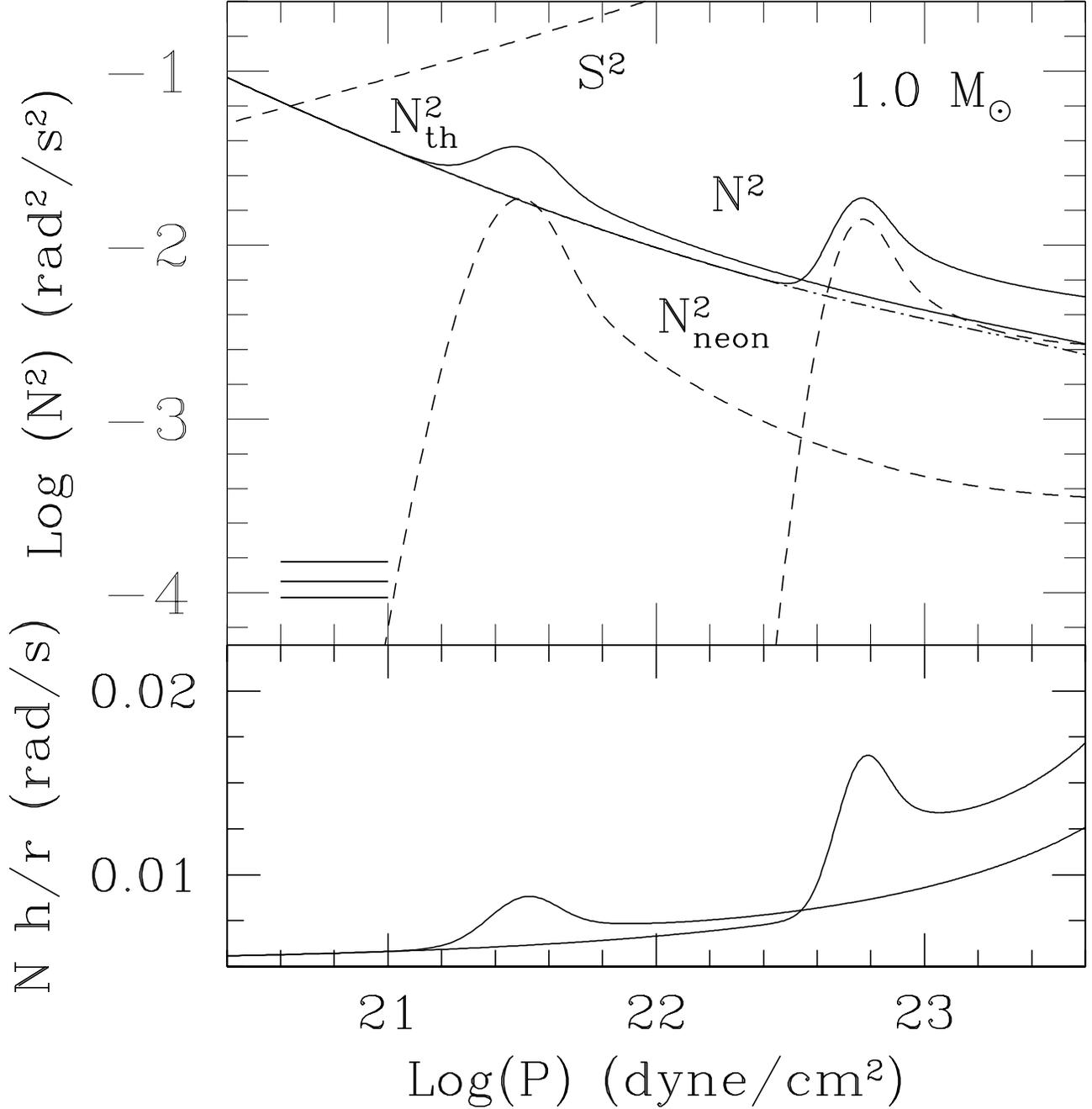}
\caption{Thermal and \xne gradient contributions to $N^2$ in a 1 $M_\odot$ WD.  Lower dashed lines give the \Ne
contribution for $D=D_s$ (leftmost curve) and $D=5D_s$ (rightmost
curve) at the time the respective model reaches $T_{\mathrm{eff}}=11900$ K. The dash-dot line is the thermal contribution, the solid curve
is the total $N^2$. The $\ell=1$ Lamb frequency is the upper dashed
line. The three horizontal lines on the left give the observed
frequencies for BPM 37093 from \citet{nitta00}. Lower Panel: The quantity $N h/r$
for the two values of $D$.
\label{fig:10brunt}}
\end{figure}

Figure \ref{fig:10brunt} shows $N^2$ for a 1 $M_\odot$ WD with
$\mu_i=14$ for $D=D_s$ and $D=5D_s$ at the point its $T_{\mathrm{eff}}=11,900$ K (The ages of the two models are 1.2 and 1.5 Gyr respectively).  For this plot, we improved upon equation (\ref{eq:n2rel}) by explicitly calculating $\chi_\rho$ using the fully degenerate equation of state instead of using either limiting approximations (see Appendix). The high pressure limit is set by the location of the crystal/fluid
boundary, which effectively acts as a hard-sphere boundary \citep{bild95,mont99b}. The solid horizontal lines show selected mode periods of the only solar mass ZZ Ceti known, BPM 37093 \citep{nitta00}.  It is clear that these observed modes probe the region of the WD interior in which \neon gradients make their largest contribution to $N^2$.

The contribution from the
\Ne abundance profile affects $N$ over the interior and thus the
periods of the longer period modes that penetrate deeply. We estimate
the extent to which \Ne sedimentation affects the mode frequencies by
performing the integral $\int N h/r d(\ln P)$ through the liquid WD
interior (the envelope's
contribution to the integral is $\approx 0.3 \ {\rm rad \ s^{-1}}$).
Using only the thermal contribution to $N^2$ gives $0.073 \ {\rm rad s^{-1}}$ through the WD core. The same core integral evaluated taking
into account the \Ne gradient gives $0.079 \ {\rm rad \ s^{-1}}$ for
$D=D_s$ and $0.095 \ {\rm rad \ s^{-1}}$ for $D=5D_s$, or a total
change in the interior integral of 8-30\%.  Adding the envelope
contribution tells us that \neon sedimentation can affect pulsation
periods by greater than 1\% , or over 100 times larger than the measurement
error. The effects of \neon sedimentation in a
$0.6 M_\odot$ WD are about 0.6\%, or 4 seconds for 700 second modes, at the levels where work is already in progress (e.g. \citet{met01, brad01}). The question of whether or not the seismological effects of \neon gradients can be separated from those due to the uncertainties in the C/O profile and envelope layer thicknesses will have to wait until detailed models are made that include the \neon contributions.  It seems probable that once this is done, it will be seen that distinguishing between theoretical mode spectra calculated with or without \neon gradient contributions is possible based on the fact that the \neon gradients introduce a new spike in the Brunt-V\"{a}is\"{a}l\"{a} frequency in the WD interior.  Whether or not pulsation studies will be able to confirm the presence of such gradients in real WDs remains to be answered.

\section{Summary and Conclusions \label{sec:conclusions}}
With two extra neutrons (over and above the $A = 2 Z$ ratio of the background ions), the diffusion of \neon is biased inward in the liquid interior of WDs. The impact of this sedimentation on WD cooling was first estimated by BH01.  We have extended their work by calculating the mutual evolution of the \neon density, $\rho_{22}$, and core temperature, $T_c$, for WD models composed of a single background ion species.  The heating produced by \neon increases the cooling age of our models on order of 0.2-1.5 Gyr (see Figure \ref{fig:delays}) depending on the value of $D$.   We have also performed initial estimates of the effects that \neon abundance gradient can have on the pulsation modes of WDs.  Although more precise work needs to be done, these initial calculations indicate that the corrections to mode periods due to $\mu_e$ gradients impact the results at the current level of observational precision.

The uncertainty in the actual value of the diffusion coefficient produces a large uncertainty in the possible impact \neon may have on WD evolution. The significant impact that \neon sedimentation may have on ages of recently formed WDs provides clear astrophysical motivation for authoritative calculations for the  $D$ of \neon through a multicomponent plasma.  Hopefully further theoretical work in this direction will eliminate this source of uncertainty.

Observations can constrain the diffusion rate of \neon in WD interiors.  For example, say the age of a WD is known because of its membership in a cluster.  Its luminosity will then depend on the rate at which \neon sinks and on the overall amount of \neon present in the WD (see equation (\ref{eq:Lg})). Also the amount by which the brightness of the WD will be affected is very dependent on the WDs mass, as can be seen by the differences between the 0.6 and 1.0 $\msun$ cooling curves in Figure \ref{fig:Ltcrys}.  Having an independent measure of these four quantities ($M$, $L$, WD age, and \xne) can allow meaningful tests of our theory.  The main uncertainty in WD cooling, namely fractionation effects, do not come into play until a large portion of the WD has crystallized and do not seem to be a function of \neon concentration (at least for $\xne \lesssim 0.1$; see \citet{seg96}).  Increases in WD luminosity over that expected from the canonical cooling theory for times prior to crystallization or which show a strong metallicity dependence are most readily interpreted as the effects of \neon sedimentation.  The question is whether or not \neon can affect the WD luminosity to a measurable degree in the regimes where fractionation is unimportant.

One place we might look are massive young WDs. For ages between 1--4 Gyrs, the $L$ of such objects is strongly dependent on $D$. But, on the other hand, these WDs begin to crystallize in this time frame. For $D \sim D_s$, C/O fractionation effects may dominate over \neon sedimentation, at least if $\xne \sim 0.02$ (\citet{sal00} and see also Figure \ref{fig:delays}). For larger values of $D$ though, these WDs will be maintained at a high enough temperature that fractionation becomes significantly less important at these times. The other place we might look for the effects of \neon are in WDs born from metal rich progenitors.  In this case, the effects on \neon sedimentation even in the lower mass WDs are more readily apparent and may be observable. 

The data required for either of the above two observational programs can be obtained for WDs that are members of open clusters.  In this case, the age of the WD can be inferred from the difference between the cluster's turn-off age and the main sequence lifetime of it progenitor, if the WD mass is known. The cluster's metallicity constrains the \neon content of the WDs. To date, though, the actual quantity of data for confirmed WD cluster members is rather scarce.  A compilation of such objects was performed by von Hippel in 1998 \citep{vonhip98}.  In this sample (which is still seemingly complete), there is only one WD (G152 in NGC 2682) with a mass or $g$ determination in a cluster older than 1 Gyr \citep{lands98}.  The other 27 objects all reside in clusters younger than this.  Of these, there are about 5-6 WDs with masses greater than 0.9 $\msun$ and several (most notably the two WD members of NGC 2168) whose mass determination are highly suspect \citep{koes88}.  Obviously, it would be ideal to increase the number of known WDs in clusters with a wider range of ages than we have currently. Such a program should be a priority if for no other reason than the need for a direct observational test of WD cooling theory in general.

In the past seven years or so, the number of WD candidates in open clusters have grown substantially due to the series of deep (limiting magnitudes of 24-26 in the $V$ band) photometric studies of open clusters that several groups have undertaken \citep{vonhip95, rich98, kali2, kali3, vonhip00,vonhip98b,andr02}.  The goal of these studies is to provide a determination of cluster ages using the age of the oldest WD candidates found in each cluster.  Identification of an object as a WD is based on the object's location in the cluster's color-magnitude diagram and on it having a spatial morphology that is stellar in nature (as opposed to one that is galactic---contamination of the WD sequence with faint blue galaxies in these studies is one of the  difficulties encountered)\citep{vonhip00, kali2}.  To date, no follow up spectroscopy to confirm these objects aras WDs and to make mass determinations if they are has been done, although the CFHT group has future plans to do so \citep{kali3}.  Overall the number of new WD candidates added by these studies is around 300 objects.  The clusters studied so far range in age from 0.5-7 Gyrs.  It is worth noting that only one of the 19 clusters in the CFHT survey has an age greater than 1 Gyr \citep{kali1}.  Follow up spectroscopy on clusters outside of the CFHT sample is thus highly desirable.  As mentioned earlier, the cluster NGC 6791 is an extremely interesting target due to its high metal content.  To our knowledge, there have been no WD candidates identified in this cluster and discovery of a WD sequence in NGC 6791 could provide a ready test of our theory since the effects of \neon on WD cooling will be greatly amplified there.

We would like to thank Peter H\"{o}eflich for providing the pre-WD evolutionary models from which we based our discussion of WD composition, Leandro Althaus for providing the $L-T_c$ relations used in our calculations, and James Liebert for a critical reading of our manuscript. This work was supported by the NSF under Grants PHY99-07949, AST01-96422, and AST02-05956.  L. B. is a Cottrell Scholar of the Research Corporation. 

\appendix
\section{Approximation of the  Brunt-V\"{a}is\"{a}l\"{a} Frequency}
The Brunt-V\"{a}is\"{a}l\"{a} frequency is given by the expression 
\begin{equation}
N^2 = -g \left[ \der{\ln \rho}{z} - \frac{1}{\Gamma_1} \der{\ln P}{z}\,, \right] 
\end{equation}
which can be expressed in the form
\begin{equation} \label{eq:bv2}
N^2 = -g \left[ \frac{1}{\chi_{\rho}} \der{\ln P}{z} - \frac{1}{\Gamma_1} \der{\ln P}{z} \right] + g \left(\frac{\chi_T}{\chi_{\rho}} \der{\ln T}{z} + \frac{\chi_{\mu_e}}{\chi_{\rho}} \der{\ln \mu_e}{z}  \right) 
\end{equation}
where
\begin{equation}
\chi_Q = \pder{\ln P}{\ln Q}
\end{equation}
and $\mu_e$ is the electron mean molecular weight.

As $\Gamma_1$ can be written 
\begin{equation} 
\Gamma_1 = \frac{\chi_\rho}{1- \nabla_{ad} \chi_T}\;,
\end{equation}
and putting $\nabla = d \ln T/d \ln P$ along with using hydrostatic balance, $d P/d z = -\rho g$, equation (\ref{eq:bv2}) can be expressed in the form
\begin{equation}
N^2 = \frac{g^2 \rho}{P} \frac{\chi_T}{\chi_\rho}\left[\nabla_{ad}-\nabla-\frac{\chi_{\mu_e}}{\chi_T} \der{\ln \mu_e}{\ln P}\right]\;,
\end{equation}
showing the explicit relation between our expression and the more common form of equation (13) from \citet{bras91}.

In extremely degenerate matter, $\,\chi_\rho \approx \Gamma_1 = d \ln P / d \ln \rho \arrowvert_{S}$ and if not careful the important term is missed. We thus rewrite the expression for the adiabatic exponent in terms of $\chi_\rho$ plus a small correction term
\begin{equation}
\Gamma_1 = \chi_\rho + \chi_T \der{\ln T}{\ln \rho}\Big\arrowvert_{S}\;.
\end{equation}

After an expansion in the small quantities and presuming $d T/dz=0$, we find 
\begin{equation}
N^2 = -g  \der{\ln P}{z}\frac{\chi_T}{\chi_\rho^2} \der{\ln T}{\ln \rho}\Big\arrowvert_{S}+ g \frac{\chi_{\mu_e}}{\chi_{\rho}} \der{\ln \mu_e}{z} \;.
\end{equation}

Now for electron degenerate matter and ideal ions 
\begin{equation}
 \chi_T = \pder{\ln P}{\ln T} = \frac{T}{P_e} \pder{P_i}{T} = \frac{T}{P_e} \frac{\rho k}{\mu_i m_p} = \frac{P_i}{P_e}\;.
\end{equation}

Also, as we take the ions to be ideal we have
\begin{equation}
\pder{\ln T}{\ln \rho}\Big\arrowvert_{S} = \frac{2}{3}\;. 
\end{equation}

In degenerate matter,$\chi_\rho = -\chi_{\mu_e}$ is between 4/3 and 5/3 depending on the relativity parameter.  Finally
\begin{equation}
\mu_e = \frac{22}{11-X_{22}} = 2 (1 + X_{22}/11)
\end{equation}
where the last step is valid only if \neon is a trace.  With this expression we have
\begin{equation}
\frac{1}{\mu_e} \der{\mu_e}{z} \approx \frac{1}{11} \der{X_{22}}{z} 
\end{equation}
so that, for relativistic matter, 
\begin{equation}
\begin{split}
N^2 &\approx  g^2 \frac{\rho}{P} \frac{P_i}{P}\frac{2}{3} \frac{9}{16} - g \frac{1}{11} \der{X_{22}}{z}\\
   & =  g^2 \frac{3}{8}\frac{P_i}{P^2}\rho - g \frac{1}{11} \der{X_{22}}{z}
\end{split}
\end{equation}

\clearpage
\clearpage

\clearpage
\clearpage

\clearpage
\clearpage

\clearpage
\clearpage

\clearpage
\clearpage
\clearpage

\clearpage
\clearpage

\clearpage
\clearpage
\clearpage

\clearpage
\clearpage

\clearpage
\clearpage
\clearpage

\clearpage
\clearpage

\clearpage
\clearpage
\clearpage

\end{document}